\documentclass[11pt,a4paper]{article}

\usepackage{amssymb}

\usepackage[dvips]{graphicx}

\begin{document}

\title{\bf The higher derivative regularization and quantum
corrections in ${\cal N}=2$ supersymmetric theories}

\author{
I.L.Buchbinder\\
{\small {\em Department of Theoretical Physics,
Tomsk State Pedagogical University,}}\\
{\small {\em 634061, Tomsk, Russia}}\\ {\small {\em and}} \\
 {\small {\em National Research Tomsk State University, 634050, Tomsk, Russia}}
\\
\\
K.V.Stepanyantz\\
{\small{\em Department of Theoretical Physics, Faculty of Physics,
Moscow State University,}}\\
{\small{\em 119991, Moscow, Russia}}}

\maketitle

\begin{abstract}
We construct a new version of the higher covariant derivative
regularization for a general ${\cal N}=2$ supersymmetric gauge
theory formulated in terms of ${\cal N}=1$ superfields. This
regularization preserves both supersymmetries of the classical
action, namely, the invariance under the manifest ${\cal N}=1$
supersymmetry and under the second hidden on-shell supersymmetry.
The regularizing ${\cal N}=2$ supersymmetric higher derivative
term is found in the explicit form in terms of ${\cal N}=1$
superfields. Thus, ${\cal N}=2$ supersymmetry is broken only by
the gauge fixing procedure. Then we analyze the exact NSVZ
$\beta$-function and prove that in the considered model its higher
loop structure is determined by the anomalous dimension of the
chiral superfield $\Phi$ in the adjoint representation which is
the ${\cal N}=2$ superpartner of the gauge superfield $V$. Using
the background field method we find that this anomalous dimension
is related with the anomalous dimension of the hypermultiplet and
vanishes if the effective action is invariant under ${\cal N}=2$
background supersymmetry. As a consequence, in this case the
higher loop contributions to $\beta$-function also vanish. The
one-loop renormalization structure in the considered
regularization is also studied by the explicit calculations of the
one-loop renormalization constants.
\end{abstract}

\unitlength=1cm



\section{Introduction}
\hspace{\parindent}

Supersymmetric theories possess remarkable properties on the
quantum level. These properties are provided by
non-renormalization theorems according to which supersymmetric
theories have a much better ultraviolet behavior than the
non-supersymmetric ones. The most famous example is, certainly,
the $D=4$, ${\cal N}=4$ supersymmetric Yang--Mills (SYM) theory,
which is finite conformal invariant quantum field theory model
\cite{Grisaru:1982zh,Mandelstam:1982cb,Brink:1982pd,Howe:1983sr}.
Perturbative quantum corrections in $D=4$, ${\cal N}=2$ SYM
theories are finite starting from the two-loop approximation
\cite{Howe:1983wj, Howe:1983sr, Buchbinder:1998}. $D=4$, ${\cal
N}=1$ supersymmetric theories have less number of independent
renormalization constants in comparison with the
non-supersymmetric ones and the superpotential has no quantum
corrections \cite{Grisaru:1979wc}. Moreover, it is possible to
find an expression for a $\beta$-function of ${\cal N}=1$
supersymmetric theories, which is exact in all orders
\cite{Novikov:1983uc,Jones,Novikov:1985rd,Shifman:1986zi,Vainshtein:1986ja,Shifman:1985fi}.
(For theories containing chiral matter superfields this expression
relates the $\beta$-function with the anomalous dimension of the
matter superfields.) This expression for the $\beta$-function is
called the exact Novikov, Shifman, Vainshtein, and Zakharov (NSVZ)
$\beta$-function.

The most elegant approach to the non-renormalization theorems is
obtained in the framework of the superfield formulation of
supersymmetric theories. In this case, the proof of the
non-renormalization theorems is based on three points: (i)
superspace structure of superpropagators, containing the
delta-functions of anticommuting variables that allow to shrink
the loops into dots in $\theta$ space, (ii) the superfield
background field method for supersymmetric gauge theories, and
(iii) an assumption about existence of a regularization manifestly
preserving the supersymmetry (see e.g. \cite{Gates:1983nr,
West:1990tg, Buchbinder:1998qv}). First two points are realized in
the explicit form. As to the last one, it is not very clear from
the beginning, how to construct a regularization which preserves
the supersymmetry and  which will be convenient for practical
computations of supergraphs (see e.g. \cite{Jack:1997sr}).
Therefore, a part of the proofs of the non-renormalization
theorems based on the above assumption needs an additional
justification.

It is known that the usually used dimensional regularization
\cite{'tHooft:1972fi,Bollini:1972ui,Ashmore:1972uj,Cicuta:1972jf},
explicitly breaks the supersymmetry (see e.g.
\cite{Delbourgo:1974az}), because numbers of bosonic and fermionic
degrees of freedom differently depend on the space-time dimension.
Most calculations of quantum corrections in supersymmetric
theories are done using the regularization by the dimensional
reduction \cite{Siegel:1979wq}, which is a special modification of
the dimensional regularization, and the
$\overline{\mbox{DR}}$-scheme. Using $\overline{\mbox{DR}}$-scheme
the finiteness of the ${\cal N}=4$ SYM theory was verified by
explicit calculations in one- \cite{Ferrara:1974pu}, two-
\cite{Jones:1977zr,Poggio:1977ma}, three-
\cite{Avdeev:1980bh,Grisaru:1980nk,Caswell:1980ru}, and four-loop
\cite{Velizhanin:2010vw} approximations. Vanishing of two- and
three-loop contributions to the $\beta$-function of the ${\cal
N}=2$ SYM theory was explicitly demonstrated in
\cite{Avdeev:1981ew}. The $\beta$-function for a general ${\cal
N}=1$ SYM theory was calculated in one- \cite{Ferrara:1974pu},
two- \cite{Jones:1974pg}, three-
\cite{Avdeev:1981ew,Jack:1996vg,Jack:1996cn}, and four-loop
\cite{Harlander:2006xq} approximations. (The result agrees with
the exact NSVZ $\beta$-function only in the one- and two-loop
approximations, where a $\beta$-function is scheme-independent. In
the higher orders the NSVZ $\beta$-function can be obtained only
after a specially constructed finite renormalization
\cite{Jack:1996vg}.)

However, it is known that the dimensional reduction is not
consistent from the mathematical point of view
\cite{Siegel:1980qs}. Due to this inconsistency any ${\cal N}$
supersymmetry can be broken by quantum corrections in higher loops
\cite{Avdeev:1982np,Avdeev:1982xy}. This means that the
non-renormalization theorems are not completely justified in
framework of dimensional reduction and the problem of their
justification is in general open.

Other methods can be also used \cite{Shifman:1985tj,Mas:2002xh}
for calculations of quantum corrections. In principle, it is
possible even to use non-invariant regularizations, if a
subtraction scheme is tuned in such a way that the Slavnov--Taylor
identities are valid for the renormalized effective action
\cite{Collins:1984xc,Slavnov:2001pu,Slavnov:2002ir,Slavnov:2002kg,Slavnov:2003cx}.
However, for practical purposes it is much better to use an
invariant regularization. Moreover, the existence of a
regularization which preserves supersymmetries of a theory is a
key step for proving the non-renormalization theorems. In order to
construct an invariant regularization it is convenient to
formulate a theory in terms of ${\cal N}=1$ superfields, because
in this case ${\cal N}=1$ supersymmetry is a manifest symmetry. A
mathematically consistent invariant regularization, which does not
break ${\cal N}=1$ supersymmetry, is the higher covariant
derivative regularization \cite{Slavnov:1971aw,Slavnov:1972sq}. In
the supersymmetric case it can be formulated in terms of ${\cal
N}=1$ superfields \cite{Krivoshchekov:1978xg,West:1985jx} and,
therefore, does not break ${\cal N}=1$ supersymmetry.

In addition to the supersymmetric regularization, manifestly
supersymmetric quantization of a theory also requires
supersymmetric gauge fixing procedure. In the case of ${\cal N}=1$
supersymmetric theories structure of divergences can be studied
either using the component fields in the Wess--Zumino gauge, or
using the superfield formulation. In the latter case the gauge can
be fixed without breaking ${\cal N}=1$ supersymmetry
\cite{Slavnov:1974uv} and the quantum corrections are calculated
in a manifestly ${\cal N}=1$ supersymmetric way that makes this
procedure very convenient. Application of the higher covariant
derivative regularization (complemented by a supersymmetric gauge
condition) to calculation of quantum corrections for ${\cal N}=1$
supersymmetric theories allows to explain naturally the origin of
the exact NSVZ $\beta$-function. Loop integrals for the
$\beta$-function appear to be integrals of total derivatives
\cite{Soloshenko:2003nc,Pimenov:2009hv,Stepanyantz:2011zz} and
even integrals of double total derivatives
\cite{Smilga:2004zr,Stepanyantz:2011bz,Stepanyantz:2012zz,
Stepanyantz:2012us,Stepanyantz:2011jy}. A qualitative explanation
of this fact can be given by analyzing Feynman rules
\cite{Stepanyantz:2011wq} using a method proposed in
\cite{Stepanyantz:2005wk}. Because the integrands in integrals
which determine a $\beta$-function are total derivatives, at least
one of the loop integrals can be calculated analytically. This
gives the NSVZ relation between the $\beta$-function and the
anomalous dimension which are defined in terms of the bare
coupling constant \cite{Kataev:2013eta}. For the ${\cal N}=1$
supersymmetric electrodynamics this was proved in all orders
\cite{Stepanyantz:2011jy}. As a consequence, in the Abelian case
the NSVZ $\beta$-function was obtained exactly in all orders of
the perturbation theory for the renormalization group functions
defined in terms of the bare coupling constant. If the
renormalization group functions are defined in terms of the
renormalized coupling constant, the NSVZ $\beta$-function is
obtained in a special subtraction scheme, which can be naturally
constructed if the theory is regularized by higher covariant
derivatives \cite{Kataev:2013eta,Kataev:2013csa}. This (NSVZ)
scheme is obtained by imposing the boundary conditions
(\ref{NSVZ_Scheme}) on the renormalization constants. However, so
far there is no proof that in non-Abelian theories the exact NSVZ
$\beta$-function is obtained with the higher covariant derivative
regularization in all orders. Nevertheless, arguments based on
anomalies \cite{Shifman:1986zi} and explicit calculations in the
lowest loops
\cite{Pimenov:2009hv,Stepanyantz:2011zz,Stepanyantz:2011bz,
Stepanyantz:2012zz,Stepanyantz:2012us} allow to suggest this.
Therefore, using the invariant regularization for ${\cal N}=1$
supersymmetric theories it is possible to make general conclusions
concerning the structure of divergences.

Existence of an invariant regularization is also needed for
proving that the $\beta$-function of ${\cal N}=2$ SYM theories
vanishes beyond the one-loop approximation
\cite{Howe:1983wj,Howe:1983sr,Buchbinder:1998}. The higher
covariant derivative regularization is formulated in terms of
${\cal N}=1$ superfields. Certainly, ${\cal N}=2$ SYM theories can
be written in terms of ${\cal N}=1$ superfields (see e.g.
\cite{Gates:1983nr}). However, in this case only ${\cal N}=1$
supersymmetry is manifest, the second supersymmetry being hidden
and on-shell. Versions of the higher covariant derivative
regularization so far used for explicit calculations preserve only
${\cal N}=1$ supersymmetry. (A version of the higher derivative
regularization for ${\cal N}=2$ supersymmetric theories was
constructed in \cite{Krivoshchekov:1985pq}, but the higher
derivative term, which is invariant under both supersymmetries,
was not presented.) Therefore, the ${\cal N}=1$ higher derivative
regularization being applied to ${\cal N}$ extended supersymmetric
theories can only state that ${\cal N}=1$ supersymmetry is
preserved. It does not guarantee that total ${\cal N}=2$
supersymmetry is not broken by quantum corrections. Therefore, the
effective action is invariant only under the manifest
supersymmetry, and it is not clear, whether it is invariant under
the second (hidden) supersymmetry.

In this paper, using the formulation of ${\cal N}=2$ SYM theories
in terms of ${\cal N}=1$ superfields, we construct a manifestly
${\cal N}=1$ supersymmetric higher covariant derivative
regularization, which is also invariant under the hidden
supersymmetry. This regularization guarantees that all loop
quantum corrections are automatically manifestly ${\cal N}=1$
supersymmetric, and invariance under the hidden supersymmetry can
be broken only by the gauge fixing procedure. (This situation is
similar to the using of the Wess-Zumino gauge in ${\cal N}=1$
supersymmetric theories: supersymmetry is also broken only by the
gauge fixing procedure.) We find that if the effective action is
invariant under the background gauge transformations and
background ${\cal N}=2$ supersymmetry, then all anomalous
dimensions of the chiral superfields vanish. Staring from the
exact NSVZ $\beta$-function we prove that beyond one loop the
divergences are completely determined by the anomalous dimension
$\gamma_\Phi$ of the chiral superfield $\Phi$ (which forms the
${\cal N}=2$ vector supermultiplet together with the gauge
superfield $V$). As a result, vanishing the $\beta$-function
beyond the one-loop approximation depends on whether the this
anomalous dimension vanishes or not.

This paper is organized as follows: In Sect.
\ref{Section_Regularization} we construct the higher derivative
regularization for ${\cal N}=2$ supersymmetric theories which is
manifestly ${\cal N}=1$ supersymmetric and also possesses the
additional hidden supersymmetry. A higher derivative term and the
Pauli--Villars determinants proposed in this section are invariant
under both supersymmetries. (A derivation of the ${\cal N}=2$
supersymmetric higher derivative term by the Noether method is
described in Appendix \ref{Appendix_Noether}.) However, the second
supersymmetry is broken by the gauge fixing procedure. The
renormalization of the considered model is discussed in Sect.
\ref{Section_Renormalization}. In Sect. \ref{Section_Invariance}
we prove that if the effective action is invariant under the
background ${\cal N}=2$ supersymmetry, then the anomalous
dimension $\gamma_{\Phi}$ vanishes. In Sect.
\ref{Section_Finiteness_And_NSVZ} starting from the exact NSVZ
$\beta$-function we derive a relation between $\beta$-function and
the function $\gamma_{\Phi}$. According to this relation, the NSVZ
$\beta$-function gets no corrections beyond one-loop if
$\gamma_{\Phi}$ vanishes. One-loop calculation of quantum
corrections with the constructed regularization is presented in
Sect. \ref{Section_One-Loop}. The results are briefly discussed in
the Conclusion.

\section{The higher covariant derivative regularization for
${\cal N}=2$ supersymmetric theories.}
\hspace{\parindent}\label{Section_Regularization}

In this paper we consider the ${\cal N}=2$ SYM theory with matter.
It is convenient to describe this theory in terms of ${\cal N}=1$
superfields \cite{West:1990tg,Gates:1983nr,Buchbinder:1998qv}.
Using this notation the action can be written as

\begin{eqnarray}\label{N=2_Action}
&& S = \frac{1}{2 e_0^2} \mbox{tr} \Big( \mbox{Re} \int d^4x\,
d^2\theta\, W^a W_a + \int d^4x\, d^4\theta\, \Phi^+ e^{2V} \Phi\,
e^{-2V} \Big) + \frac{1}{4} \int d^4x\,d^4\theta\,\Big(\phi^+
e^{2V} \phi \nonumber\\
&& + \widetilde\phi^+ e^{-2V^t} \widetilde\phi\Big) +
\Big(\frac{i}{\sqrt{2}} \int d^4x\,d^2\theta\, \widetilde\phi^t
\Phi \phi + \frac{1}{2} m_0 \int d^4x\,d^2\theta\,
\widetilde\phi^t \phi + \mbox{c.c.} \Big),
\end{eqnarray}

\noindent where $e_0$ is a bare coupling constant, and the real
superfield $V$ contains the gauge field $A_\mu$ as a component. A
superfield strength of the gauge superfield $V$ is defined by

\begin{equation}
W_a = \frac{1}{8} \bar{D}^2 (e^{-2V} D_a e^{2V}).
\end{equation}

\noindent (In our notation indices of right spinors are denoted by
the Latin letters, and indices of left spinors are denoted by
Latin letters with dots. Vector indices are denoted by Greek
letters.) The chiral superfield $\Phi$ belongs to the adjoint
representation of the gauge group. Together with the superfield
$V$ it forms the multiplet of the ${\cal N}=2$ SYM theory. The
chiral superfields $\phi$ and $\widetilde\phi$ form an ${\cal
N}=2$ hypermultiplet. The superfield $\phi$ lies in a
representation $R_0$, which can be, in general, reducible. The
superfield $\widetilde\phi$ lies in the conjugated representation
$\overline{R}_0$. For simplicity the action is written for a
theory with a single coupling constant (i.e. the gauge group is
simple) and a single mass $m_0$ (this corresponds to the
irreducible representation $R_0$). The results described below can
be easily generalized to more complicated cases.

The theory (\ref{N=2_Action}) is invariant under the
supersymmetric gauge transformations

\begin{eqnarray}\label{Gauge_Transformations}
&& e^{2V} \to e^{-A^+} e^{2V} e^{-A}; \qquad W_a \to e^{A} W_a
e^{-A};\qquad \Phi \to e^A \Phi e^{-A};\qquad\nonumber\\
&&\qquad\qquad\qquad\quad \phi \to e^A \phi;\qquad\qquad
\widetilde\phi \to e^{-A^t} \widetilde\phi,\vphantom{\Big(}
\end{eqnarray}

\noindent where the parameter $A$ is an arbitrary chiral scalar
superfield which takes values in the Lie algebra of the gauge
group. (In the second string $A$ should be certainly presented as
$\mbox{dim}\, R_0 \times \mbox{dim}\, R_0$ matrix.) Also the
theory (\ref{N=2_Action}) is invariant under two supersymmetries.
The first supersymmetry is a manifest symmetry, because the action
is written in terms of the ${\cal N}=1$ superfields. The
transformations of this supersymmetry can be also written in terms
of the ${\cal N}=1$ superfields \cite{Gates:1983nr}. For this
purpose it is convenient to present the exponent of the gauge
superfield as

\begin{equation}
e^{2V} = e^{\Omega^+} e^\Omega
\end{equation}

\noindent  and define the right and left spinor gauge covariant
derivatives

\begin{equation}
\nabla_a = e^{-\Omega^+} D_a e^{\Omega^+};\qquad
\bar{\nabla}_{\dot a} = e^\Omega \bar D_{\dot a} e^{-\Omega},
\end{equation}

\noindent respectively. Then the transformations of the manifest
supersymmetry in terms of the ${\cal N}=1$ superfields can be
written as

\begin{eqnarray}\label{First_SUSY}
&&\delta e^{\Omega} = - 8iD^a\xi e^\Omega W_a;\qquad \delta
e^{\Omega^+} = 8i\bar{D}^{\dot a}\xi \bar{W}_{\dot a}
e^{\Omega^+};
\vphantom{\frac{1}{2}}\nonumber\\
&& \delta \Phi = i\bar D^2 \Big[e^{-2V} D^a (e^{2V}\Phi e^{-2V})
e^{2V} D_a\xi\Big];\vphantom{\frac{1}{2}}\nonumber\\
&& \delta\phi = i \bar D^2\Big(e^{-2V} D^a \xi\, D_a(e^{2V}\phi)
+ \frac{1}{2} D^2\xi \phi\Big);\nonumber\\
&& \delta\widetilde \phi = i \bar D^2\Big(e^{2V^t} D^a \xi\,
D_a(e^{-2V^t}\widetilde \phi) + \frac{1}{2} D^2\xi \widetilde
\phi\Big),
\end{eqnarray}

\noindent where $\xi$ is a real scalar superfield which does not
depend on the space-time coordinates. This superfield is a
parameter of the transformations of the manifest supersymmetry.
The action (\ref{N=2_Action}) is also invariant under the
transformations of the second on-shell supersymmetry. In terms of
the ${\cal N}=1$ superfields these transformations have the form

\begin{eqnarray}\label{Second_SUSY}
&& \delta e^\Omega = i\eta^* e^\Omega \Phi;\qquad \delta
e^{\Omega^+} = -i\eta \Phi^+ e^{\Omega^+};\qquad \delta\Phi =
-\frac{i}{2} W^a D_a\eta;\nonumber\\
&& \delta\phi = -\frac{1}{4\sqrt{2}} \Big(\bar D^2 (\eta^* e^{-2V}
\widetilde\phi^*) -4m_0 \eta \phi\Big);\qquad \delta\widetilde
\phi = \frac{1}{4\sqrt{2}} \Big(\bar D^2 (\eta^* e^{2V^t} \phi^*)
-4m_0 \eta \widetilde\phi\Big),\qquad
\end{eqnarray}

\noindent where $\eta$ is a chiral superfield independent of the
space-time coordinates.

In order to regularize the theory (\ref{N=2_Action}) we add to its
action a term $S_\Lambda$ with higher covariant derivatives.
Certainly, this term is not uniquely defined, because a number of
derivatives can be arbitrary. In this paper we construct the
simplest variant of this term, which is proportional to
$\Lambda^{-2}$, where $\Lambda$ is a dimensionful parameter with
the dimension of a mass. In Appendix \ref{Appendix_Noether} using
the Noether method we construct an expression for the action
$S_\Lambda$ invariant under ${\cal N}=2$ supersymmetry. It can be
written in the following form:

\begin{eqnarray}\label{Higher_Derivative_Term}
&& S_\Lambda = -\frac{1}{16 e_0^2\Lambda^2} \mbox{tr} \int
d^4x\,\Bigg\{\frac{1}{2} \mbox{Re} \int d^2\theta\, (e^\Omega W^a
e^{-\Omega}) \bar\nabla^2 \nabla^2 (e^\Omega W_a
e^{-\Omega})\nonumber\\
&& + \int d^4\theta\,\Bigg(\frac{1}{2}(e^{-\Omega^+} \Phi^+
e^{\Omega^+}) \bar\nabla^2 \nabla^2 (e^\Omega \Phi e^{-\Omega}) +
4(e^\Omega W^a e^{-\Omega}) \Big[\nabla_a(e^\Omega \Phi
e^{-\Omega}),(e^{-\Omega^+} \Phi^+ e^{\Omega^+})\Big]
\vphantom{\frac{1}{2}}\nonumber\\
&& + 4(e^{-\Omega^+} \bar W^{\dot a} e^{\Omega^+}) \Big[(e^\Omega
\Phi e^{-\Omega}),\bar\nabla_{\dot a}(e^{-\Omega^+} \Phi^+
e^{\Omega^+})\Big] -8 \Big[ (e^\Omega \Phi e^{-\Omega}),
(e^{-\Omega^+} \Phi^+ e^{\Omega^+})\Big]^2 \Bigg)\Bigg\}.
\end{eqnarray}

\noindent It is easy to see that after adding this term the
divergences remain only in one-loop supergraphs, which is a
typical feature of the higher derivative regularization
\cite{Faddeev:1980be}. Therefore, it is necessary to regularize
the remaining one-loop divergences. Usually for this purpose the
Pauli--Villars determinants should be inserted into the generating
functional \cite{Slavnov:1977zf}. However, in the considered case
it is necessary to do this very carefully, because this procedure
should not break ${\cal N}=2$ supersymmetry. It is necessary to
introduce two different sets of the Pauli--Villars fields: the
first set cancels one-loop divergences originated by the ${\cal
N}=2$ gauge supermultiplet and the second one cancels one-loop
divergences originated by the hypermultiplet. Taking into account
absence of quadratic divergences for the ${\cal N}=2$ SYM theory,
in order to cancel one-loop divergences of the gauge
supermultiplet (and ghosts), it is sufficient to use a single
Pauli--Villars determinant

\begin{equation}
\mbox{Det}(PV, M)^{-1} = \int D\varphi\,D\widetilde\varphi \exp(i
S_{\varphi}),
\end{equation}

\noindent where the action for the Pauli--Villars fields $\varphi$
and $\widetilde\varphi$ (in the adjoint representation of the
gauge group) is given by

\begin{eqnarray}\label{PV_Action}
&& S_{\varphi} = \frac{1}{2 e_0^2} \mbox{tr}\int
d^4x\,d^4\theta\,\Big(\varphi^+ e^{2V} \varphi e^{-2V} +
\widetilde\varphi^+ e^{2V} \widetilde\varphi
e^{-2V}\Big)\nonumber\\
&& + \frac{1}{e_0^2} \mbox{tr}\Big(\int d^4x\,d^2\theta\, \Big(
i\sqrt{2}\, \widetilde\varphi [\Phi, \varphi] + M_0
\widetilde\varphi \varphi\Big) + \mbox{c.c.} \Big).
\end{eqnarray}

\noindent We choose the mass $M_0$ of these Pauli--Villars fields
proportional to the dimensionful parameter $\Lambda$ in the higher
derivative term:

\begin{equation}
M_0 = a_0\Lambda,
\end{equation}

\noindent where the finite constant $a_0$ does not depend on the
bare coupling constant. Introducing the Pauli--Villars fields
$\varphi$ and $\widetilde\varphi$ is motivated by the analogy with
the ${\cal N}=4$ SYM theory. Really, in the ${\cal N}=4$ SYM
theory 3 chiral superfields in the adjoint representation of the
gauge group compensate divergences from the gauge supermultiplet
and ghosts. This allows to guess that two Pauli--Villars fields
$\varphi$ and $\widetilde\varphi$ compensate at least a one-loop
divergence originated by the gauge supermultiplet and one chiral
superfield $\Phi$ in the adjoint representation (including
one-loop divergences of the ghost loop). This statement is
verified by the explicit calculation made in Sect.
\ref{Section_One-Loop}. Moreover, the action (\ref{PV_Action}) is
evidently invariant under both supersymmetries, because it
coincides with the action of the massive ${\cal N}=2$
hypermultiplet in the adjoint representation of the gauge group.

Also we insert in the generating functional the Pauli--Villars
determinants which cancel one-loop divergences originated by the
hypermultiplet:

\begin{equation}
\prod\limits_{I=1}^n \mbox{Det}(PV, M_I)^{c_I},
\end{equation}

\noindent where the coefficients $c_I$ satisfy the conditions

\begin{equation}
\sum\limits_{I=1}^n c_I =1;\qquad \sum\limits_{I=1}^n c_I M_I^2 =
0.
\end{equation}

\noindent Again, it is convenient to present the Pauli--Villars
determinants in the form

\begin{equation}
\mbox{Det}(PV, M_I)^{-1} = \int D\phi_I\,D\widetilde\phi_I \exp(i
S_{I}),
\end{equation}

\noindent where $\phi_I$ lies in the same representation $R_0$ as
the fields $\phi$,

\begin{equation}
S_I \equiv \frac{1}{4} \int d^4x\,d^4\theta\,\Big(\phi_I^+ e^{2V}
\phi_I + \widetilde\phi_I^+ e^{-2V^t} \widetilde\phi_I \Big) +
\Big(\int d^4x\,d^2\theta\, \Big(\frac{i}{\sqrt{2}}\,
\widetilde\phi_I^t \Phi \phi_I + \frac{1}{2} M_I
\widetilde\phi_I^t \phi_I\Big) + \mbox{c.c.} \Big),
\end{equation}

\noindent and the masses are proportional to the parameter
$\Lambda$:

\begin{equation}
M_I = a_I\Lambda,
\end{equation}

\noindent $a_I$ being independent of $e_0$. Both Pauli--Villars
actions are invariant under the transformations of ${\cal N}=2$
supersymmetry, because they coincide with the actions for the
massive ${\cal N}=2$ hypermultiplets. Therefore, the
regularization procedure is also invariant under both
supersymmetries.

The next step is gauge fixing. We will do this in the framework of
the ${\cal N}=1$ superfield background field method
\cite{Gates:1983nr}, which allows to get a manifestly ${\cal N}=1$
supersymmetric effective action preserving the classical gauge
invariance. The gauge superfield $V$ is split into the background
field and the quantum field by making the substitution

\begin{equation}
e^{\Omega} \to e^{\Omega_T} = e^{\Omega} e^{\mbox{\scriptsize
\boldmath$\Omega$}};\qquad e^{\Omega^+} \to e^{\Omega_T^+} =
e^{\mbox{\scriptsize \boldmath$\Omega$}^+} e^{\Omega^+}.
\end{equation}

\noindent Then the background superfield $\mbox{\boldmath$V$}$ and
the quantum superfield $V$ are defined by

\begin{equation}
e^{2\mbox{\scriptsize \boldmath$V$}} \equiv e^{\mbox{\scriptsize
\boldmath$\Omega$}^+} e^{\mbox{\scriptsize
\boldmath$\Omega$}};\qquad e^{2V} \equiv e^{\Omega^+} e^{\Omega}.
\end{equation}

\noindent The background-quantum splitting for the chiral
superfield $\Phi$ is trivial:

\begin{equation}
\Phi \to \Phi_T = \Phi + \mbox{\boldmath$\Phi$},
\end{equation}

\noindent where in the right hand side $\mbox{\boldmath$\Phi$}$ is
the background superfield and $\Phi$ is the quantum superfield.
The gauge fixing term used in this paper does not include the
superfields $\Phi$ and $\mbox{\boldmath$\Phi$}$. As a consequence,
the effective action depends only on the sum
$\Phi+\mbox{\boldmath$\Phi$}$. This can be easily verified by
making the linear substitution $\Phi \to \Phi_T$ in the generating
functional. Due to the same reason we do not use the background
field method for the other chiral matter superfields.

It is convenient to fix a gauge without breaking the background
gauge invariance

\begin{eqnarray}\label{Background_Gauge_Transformations}
&& e^{\mbox{\scriptsize \boldmath$\Omega$}} \to e^{i K}
e^{\mbox{\scriptsize \boldmath$\Omega$}} e^{-A};\qquad\
e^{\mbox{\scriptsize \boldmath$\Omega$}^+} \to e^{-A^+}
e^{\mbox{\scriptsize \boldmath$\Omega$}^+} e^{-i K};\qquad\
e^\Omega \to e^\Omega e^{-i K};\qquad e^{\Omega^+} \to e^{i K}
e^{\Omega^+};\nonumber\\
&& V \to e^{iK} V e^{-iK}; \qquad \mbox{\boldmath$\Phi$} \to e^A
\mbox{\boldmath$\Phi$} e^{-A} \qquad \Phi \to e^A \Phi e^{-A};
\qquad \phi \to e^A \phi;\qquad \widetilde\phi \to e^{-A^t}
\widetilde\phi,\vphantom{\Big(}\qquad
\end{eqnarray}

\noindent where $K$ is an arbitrary real scalar superfield and $A$
is an arbitrary chiral superfield. For this purpose we use the
background covariant derivatives which are defined by

\begin{equation}
\mbox{\boldmath$\nabla$}_a = e^{-\mbox{\scriptsize
\boldmath$\Omega$}^+} D_a e^{\mbox{\scriptsize
\boldmath$\Omega$}^+};\qquad \mbox{\boldmath$\bar{\nabla}$}_{\dot
a} = e^{\mbox{\scriptsize \boldmath$\Omega$}} \bar D_{\dot a}
e^{-{\mbox{\scriptsize \boldmath$\Omega$}}}.
\end{equation}

\noindent Note that the theory is also invariant under the quantum
gauge transformations

\begin{eqnarray}\label{Quantum_Gauge_Transformations}
&& e^\Omega \to e^\Omega e^{-\mbox{\scriptsize
\boldmath$A$}};\qquad e^{\Omega^+} \to e^{-\mbox{\scriptsize
\boldmath$A$}^+} e^{\Omega^+}; \qquad e^{\mbox{\scriptsize
\boldmath$\Omega$}} \to e^{\mbox{\scriptsize
\boldmath$\Omega$}};\qquad e^{\mbox{\scriptsize
\boldmath$\Omega$}^+}\to e^{\mbox{\scriptsize
\boldmath$\Omega$}^+};\vphantom{\Big)}\nonumber\\
&&\ e^{{\mbox{\scriptsize \boldmath$\Omega$}}} \Phi
e^{-{\mbox{\scriptsize \boldmath$\Omega$}}} \to
e^{\mbox{\scriptsize \boldmath$A$}} (e^{{\mbox{\scriptsize
\boldmath$\Omega$}}} \Phi e^{-{\mbox{\scriptsize
\boldmath$\Omega$}}}) e^{-\mbox{\scriptsize \boldmath$A$}};\qquad
e^{{\mbox{\scriptsize \boldmath$\Omega$}}} \mbox{\boldmath$\Phi$}
e^{-{\mbox{\scriptsize \boldmath$\Omega$}}} \to
e^{\mbox{\scriptsize \boldmath$A$}} (e^{{\mbox{\scriptsize
\boldmath$\Omega$}}} \mbox{\boldmath$\Phi$} e^{-{\mbox{\scriptsize
\boldmath$\Omega$}}}) e^{-\mbox{\scriptsize
\boldmath$A$}}; \vphantom{\Big)}\nonumber\\
&& \qquad\qquad\quad e^{{\mbox{\scriptsize \boldmath$\Omega$}}}
\phi \to e^{\mbox{\scriptsize \boldmath$A$}}
(e^{{\mbox{\scriptsize \boldmath$\Omega$}}} \phi);\qquad
e^{-{\mbox{\scriptsize \boldmath$\Omega$}}^t} \widetilde\phi \to
e^{-\mbox{\scriptsize \boldmath$A$}^t} (e^{-{\mbox{\scriptsize
\boldmath$\Omega$}}^t} \widetilde\phi),\vphantom{\Big(}\qquad
\end{eqnarray}

\noindent where $\mbox{\boldmath$A$}$ and $\mbox{\boldmath$A$}^+$
are arbitrary background-(anti)chiral superfields:

\begin{equation}
\mbox{\boldmath$\bar \nabla$}_{\dot a} \mbox{\boldmath$A$} =
0;\qquad \mbox{\boldmath$\nabla$}_a \mbox{\boldmath$A$}^+ = 0.
\end{equation}

\noindent The generating functional can be formally written as

\begin{equation}
Z = \int D\mu\, \mbox{Det}(PV, M_0)^{-1} \prod\limits_{I=1}^n
\mbox{Det}(PV, M_I)^{c_I} \exp\Big(iS + iS_\Lambda + i
S_{\mbox{\scriptsize sources}}\Big),
\end{equation}

\noindent where the action $S + S_\Lambda$ and the Pauli--Villars
determinants are invariant under ${\cal N}=2$ supersymmetry by
construction. Then according to standard procedure
\cite{Faddeev:1980be} we insert into the generating functional

\begin{equation}\label{Insertion}
1 = \Delta[V]\cdot \int D\mbox{\boldmath$A$}
D\mbox{\boldmath$A$}^+ \delta(\mbox{\boldmath$\bar \nabla$}^2
V^{(\mbox{\scriptsize \boldmath$A$})} - \mbox{\boldmath$f$})
\delta(\mbox{\boldmath$\nabla$}^2 V^{(\mbox{\scriptsize
\boldmath$A$})} - \mbox{\boldmath$f$}^+ ),
\end{equation}

\noindent where $\mbox{\boldmath$f$}$ and $\mbox{\boldmath$f$}^+$
are background-(anti)chiral superfields which satisfy the
conditions

\begin{equation}
\mbox{\boldmath$\bar \nabla$}_{\dot a} \mbox{\boldmath$f$} =
0;\qquad \mbox{\boldmath$\nabla$}_a \mbox{\boldmath$f$}^+ = 0.
\end{equation}

\noindent The quantum gauge superfield transformed under the
infinitesimal quantum gauge transformations is denoted by

\begin{equation}
V^{(\mbox{\boldmath$A$})} =
\frac{1}{2}\ln\Big(e^{-\mbox{\scriptsize \boldmath$A$}^+} e^{2V}
e^{-\mbox{\scriptsize \boldmath$A$}}\Big) \approx V +
\Big(\frac{V}{1-e^{2V}}\Big)_{Adj} \mbox{\boldmath$A$}^+ -
\Big(\frac{V}{1-e^{-2V}}\Big)_{Adj} \mbox{\boldmath$A$}.
\end{equation}

\noindent The generating functional obtained after this insertion
of $1$ is defined by $Z[j,\mbox{\boldmath$f$}]$. It is evidently
equal to the original generating functional. Then we perform the
integration

\begin{eqnarray}
&& Z[j] \to \int D\mbox{\boldmath$f$} D\mbox{\boldmath$f$}^+
D\mbox{\boldmath$C$} D\mbox{\boldmath$C$}^+\,
Z[j,\mbox{\boldmath$f$}]\, \exp\Big(-\frac{i}{16 e_0^2} \mbox{tr}
\int d^4x\,d^4\theta\, \mbox{\boldmath$f$}^+
\Big(1-\frac{\mbox{\boldmath$\bar
\nabla$}^2\mbox{\boldmath$\nabla$}^2}{16\Lambda^2}\Big)
\mbox{\boldmath$f$}
\Big)\nonumber\\
&& \times \exp\Big(-\frac{i}{16 e_0^2}\mbox{tr}\int
d^4x\,d^4\theta\,\mbox{\boldmath$C$}^+
\Big(1-\frac{\mbox{\boldmath$\bar
\nabla$}^2\mbox{\boldmath$\nabla$}^2}{16\Lambda^2}\Big)
\mbox{\boldmath$C$} \Big).\qquad
\end{eqnarray}

\noindent The anticommuting background (anti)chiral
Nielsen--Kallosh ghosts $\mbox{\boldmath$C$}$ and
$\mbox{\boldmath$C$}^+$ (in the adjoint representation of the
gauge group) can be expressed in terms of the (anti)chiral
superfields $C$ and $C^+$ as

\begin{equation}\label{Background_Chiral_Fields}
\mbox{\boldmath$C$} = e^{\mbox{\scriptsize \boldmath$\Omega$}} C
e^{-\mbox{\scriptsize \boldmath$\Omega$}};\qquad
\mbox{\boldmath$C$}^+ = e^{-\mbox{\scriptsize
\boldmath$\Omega$}^+} C^+ e^{\mbox{\scriptsize
\boldmath$\Omega$}^+}.
\end{equation}

\noindent The integration over $C$ and $C^+$ cancels the
determinant appearing after the integration over the fields $f$
and $f^+$, which are defined using equations similar to
(\ref{Background_Chiral_Fields}). It is convenient to present the
corresponding contribution in the form

\begin{equation}
\int DC\,DC^+ \exp\left(iS_C\right),
\end{equation}

\noindent where

\begin{equation}
S_C = -\frac{1}{16 e_0^2} \mbox{tr} \int
d^4x\,d^4\theta\,\mbox{\boldmath$C$}^+
\Big(1-\frac{\mbox{\boldmath$\bar\nabla$}^2
\mbox{\boldmath$\nabla$}^2}{16\Lambda^2}\Big) \mbox{\boldmath$C$}
\end{equation}

\noindent is the action for the Nielsen--Kallosh ghosts. (It is
assumed that the fields $\mbox{\boldmath$C$}$ and
$\mbox{\boldmath$C$}^+$ are expressed in terms of $C$ and $C^+$
using Eq. (\ref{Background_Chiral_Fields}).)

Substituting the explicit expression for the functional
$Z[j,\mbox{\boldmath$f$}]$ and taking the integrals over
$\mbox{\boldmath$f$}$ and $\mbox{\boldmath$f$}^+$ we obtain that
the gauge fixing term

\begin{equation}\label{Gauge}
S_{\mbox{\scriptsize gf}} = - \frac{1}{16 e_0} \mbox{tr} \int
d^4x\,d^4\theta\,\mbox{\boldmath$\nabla$}^2 V
\Big(1-\frac{\mbox{\boldmath$\bar\nabla$}^2
\mbox{\boldmath$\nabla$}^2}{16\Lambda^2}\Big)
\mbox{\boldmath$\bar{\nabla}$}^2 V
\end{equation}

\noindent is effectively added to the classical action
(\ref{N=2_Action}). As usually, $\Delta[V]$ is presented as an
integral over the Faddeev--Popov ghost fields and gives the ghost
action:

\begin{equation}
\Delta[V] = \int D\bar c\, Dc\, D\bar c^+ Dc^+
\exp\left(iS_{\mbox{\scriptsize ghost}}\right)
\end{equation}

\noindent where

\begin{equation}\label{Ghost_Action}
S_{\mbox{\scriptsize ghost}} = \frac{1}{e_0^2}\mbox{tr}\int
d^4x\,d^4\theta\,(\mbox{\boldmath$\bar c$}+ \mbox{\boldmath$\bar
c$}^+)\Big[ \Big(\frac{V}{1-e^{2V}}\Big)_{Adj}
\mbox{\boldmath$c$}^+ - \Big(\frac{V}{1-e^{-2V}}\Big)_{Adj}
\mbox{\boldmath$c$} \Big]
\end{equation}

\noindent with

\begin{equation}
\Big(f_0 + f_1 V + f_2 V^2 +\ldots\Big)_{Adj} X \equiv f_0 X + f_1
[V,X] + f_2 [V, [V, X]] + \ldots
\end{equation}

\noindent The ghost $\mbox{\boldmath$c$}$ and the antighost
$\mbox{\boldmath$\bar c$}$ are background-chiral; the ghost
$\mbox{\boldmath$c$}^+$ and the antighost $\mbox{\boldmath$\bar
c$}^+$ are background-antichiral. The ghost fields can be
expressed in terms of the (anti)chiral fields $c$, $c^+$, $\bar
c$, and $\bar c^+$ using equations similar to
(\ref{Background_Chiral_Fields}).

Thus, the generating functional can be written as

\begin{equation}
Z = \int D\mu\, \mbox{Det}(PV, M_0)^{-1} \prod\limits_{I=1}^n
\mbox{Det}(PV, M_I)^{c_I} \exp\Big(iS + iS_\Lambda +
S_{\mbox{\scriptsize gf}} + S_{\mbox{\scriptsize ghost}} + iS_C +
i S_{\mbox{\scriptsize sources}}\Big),
\end{equation}

\noindent where $d\mu$ denotes the integration measure, $S$ is the
original action of a ${\cal N}=2$ supersymmetric theory (which can
also contain hypermultiplet superfields), $S_\Lambda$ is the
regularizing action constructed in this paper, and
$S_{\mbox{\scriptsize sources}}$ is the action for the sources.
The higher derivative term $S_\Lambda$ and the Pauli--Villars
determinants are invariant under the transformations of both
supersymmetries. However, the gauge fixing term and the ghost
action are invariant only under the transformations of the
manifest supersymmetry. Therefore, ${\cal N}=2$ supersymmetry is
broken only by the gauge fixing procedure.

\section{Renormalization}
\hspace{\parindent}\label{Section_Renormalization}

The results of the previous section show that all ingredients for
proving the ${\cal N}=1$ non-renormalization theorem take place in
the considered theory. The theory is renormalizable and
renormalization preserves the manifest ${\cal N}=1$ supersymmetry.
Therefore, the divergences can be absorbed into the redefinitions
of the coupling constant, fields, and masses:

\begin{eqnarray}\label{Renormalization}
&& \frac{1}{e_0^2} = \frac{Z_3}{e^2};\qquad m_0 = Z_m m;\qquad
\Phi_T = \sqrt{Z_\Phi} \Phi_{TR};\qquad V = Z_V V_R;\nonumber\\
&& \phi = \sqrt{Z_\phi} \phi_R;\qquad\ \widetilde\phi =
\sqrt{Z_\phi} \widetilde\phi_R; \qquad\ c =Z_c c_R;\qquad \ \bar c
= Z_{\bar c} \bar c_R.\qquad
\end{eqnarray}

\noindent Here we took into account that (as we already mentioned
above) the effective action depends only on $\Phi_T =
\Phi+\mbox{\boldmath$\Phi$}$, and, therefore, it is not necessary
to introduce two different renormalization constants for the
background and quantum parts of this field.

According to the standard prescription, the renormalization
constants $Z$ should be constructed so that to cancel the
divergences, appearing in loop integrals. $\Omega_R$ can be
defined by the equation

\begin{equation}\label{Quantum_Field_Renormalization}
e^{2V_R} = e^{\Omega_R^+} e^{\Omega_R}.
\end{equation}

\noindent It is important that in Eqs. (\ref{Renormalization}) and
(\ref{Quantum_Field_Renormalization}) $V$ denotes the {\it
quantum} gauge superfield. The {\it background} gauge superfield
$\mbox{\boldmath$V$}$ is not renormalized due to the unbroken
background gauge invariance
(\ref{Background_Gauge_Transformations}). Therefore, after the
renormalization the total gauge superfield is renormalized as

\begin{equation}\label{Total_Gauge_Superfield}
e^{2V_T} \equiv e^{\Omega_T^+} e^{\Omega_T} \equiv
e^{\mbox{\scriptsize \boldmath$\Omega$}^+} e^{2V}
e^{\mbox{\scriptsize \boldmath$\Omega$}} = e^{\mbox{\scriptsize
\boldmath$\Omega$}^+} e^{2Z_V V_R} e^{\mbox{\scriptsize
\boldmath$\Omega$}}.
\end{equation}

\noindent In our notation the renormalized fields in the adjoint
representation are presented in the form

\begin{equation}
\mbox{\boldmath$V$} = \mbox{\boldmath$V$}_R = e
(\mbox{\boldmath$V$}_R)_A t^A;\qquad V_R = e (V_R)_A t^A;\qquad
\mbox{\boldmath$\Phi$}_{R} = e (\mbox{\boldmath$\Phi$}_{R})_A t^A;
\qquad \Phi_{R} = e (\Phi_{R})_A t^A,
\end{equation}

\noindent where $e$ is the renormalized coupling constant, and
$t^A$ denotes the generators of the fundamental representation,
which are normalized by the condition

\begin{equation}
\mbox{tr}(t^A t^B) = \frac{1}{2} \delta^{AB}.
\end{equation}

\noindent The similar equations for the bare superfields have the
form

\begin{equation}\label{Bare_Adjoint_Superfields}
\mbox{\boldmath$V$} = e_0 \mbox{\boldmath$V$}_A t^A;\qquad V = e_0
V_A t^A;\qquad \mbox{\boldmath$\Phi$} = e_0
(\mbox{\boldmath$\Phi$})_A t^A;\qquad \Phi = e_0 \Phi_A t^A,
\end{equation}

\noindent where $e_0$ is the bare coupling constant. Therefore, in
components

\begin{eqnarray}
&& \mbox{\boldmath$V$}_A = \sqrt{Z_3} (\mbox{\boldmath$V$}_R)_A;
\qquad\qquad\qquad\qquad\quad\ V_A = Z_V \sqrt{Z_3} (V_R)_A;
\vphantom{\Big(}\nonumber\\
&& \mbox{\boldmath$\Phi$}_A = \sqrt{Z_\Phi Z_3}
(\mbox{\boldmath$\Phi$}_R)_A \equiv Z_A{}^B
(\mbox{\boldmath$\Phi$}_R)_B; \qquad \Phi_A = \sqrt{Z_\Phi Z_3}
(\Phi_R)_A = Z_A{}^B (\Phi_R)_B,\qquad
\end{eqnarray}

\noindent where we have introduced the notation

\begin{equation}\label{ZAB}
Z_A{}^B \equiv \sqrt{Z_\Phi Z_3} \delta_A{}^B.
\end{equation}

\noindent After substitution (\ref{Renormalization})

\begin{eqnarray}\label{Renormalized_N=2_Action}
&& S = \frac{Z_3}{2 e^2} \mbox{tr} \int d^4x\,\Big(\mbox{Re} \int
d^2\theta\, W^a W_a + Z_\Phi \int d^4\theta\,\Phi_{TR}^+ e^{2 V_T}
\Phi_{TR}\, e^{-2 V_T} \Big)\nonumber\\
&& + \frac{Z_\phi}{4} \int d^4x\,d^4\theta\,\Big(\phi_R^+ e^{2
V_T} \phi_R + \widetilde\phi_R^+ e^{-2 V_T^t}
\widetilde\phi_R\Big) + Z_\phi \Big(\frac{i
Z_\Phi^{1/2}}{\sqrt{2}} \int d^4x\,d^2\theta\,
\widetilde\phi_R^t \Phi_{TR} \phi_R\nonumber\\
&& + \frac{1}{2} Z_m m \int d^4x\,d^2\theta\, \widetilde\phi_R^t
\phi_R + \mbox{c.c.}\Big) \vphantom{\Bigg)},
\end{eqnarray}

\noindent where $W_a$ is constructed from the total gauge
superfield $V_T$ given by Eq. (\ref{Total_Gauge_Superfield}). The
quantum gauge superfield is present in the action only in the
combination $e^{2V} = e^{\Omega^+} e^\Omega$. This quantum gauge
superfield should be substituted by

\begin{equation}\label{Renormalized_V}
V = Z_V V_R.
\end{equation}

The ghost Lagrangian is renormalized as

\begin{equation}
S_{\mbox{\scriptsize ghost}} = \frac{Z_3 Z_c Z_{\bar
c}}{e^2}\mbox{tr}\int d^4x\,d^4\theta\, (\mbox{\boldmath$\bar
c$}_R + \mbox{\boldmath$\bar c$}_R^+) \Big[
\Big(\frac{V}{1-e^{2V}}\Big)_{Adj} \mbox{\boldmath$c$}_R^+ -
\Big(\frac{V}{1-e^{-2V}}\Big)_{Adj} \mbox{\boldmath$c$}_R \Big],
\end{equation}

\noindent where $V$ should be also substituted by $Z_V V_R$
according to Eq. (\ref{Renormalized_V}).

Using the ${\cal N}=1$ nonrenormalization theorem
\cite{Grisaru:1979wc} according to which the superpotential

\begin{equation}
\frac{i}{\sqrt{2}} \int d^4x\,d^2\theta\, \widetilde\phi^t \Phi
\phi + \frac{1}{2} m_0 \int d^4x\,d^2\theta\, \widetilde\phi^t
\phi
\end{equation}

\noindent is not renormalized, from Eq.
(\ref{Renormalized_N=2_Action}) we immediately obtain

\begin{equation}\label{Z_Hypermultiplet}
Z_\Phi^{1/2} = Z_m = Z_\phi^{-1}.
\end{equation}

\noindent Certainly, finite renormalizations are possible, but in
this paper we assume that for finite terms the renormalization
constants are chosen equal to 1.

\section{Counterterms and the second ${\cal N}=1$ supersymmetry}
\hspace{\parindent}\label{Section_Invariance}

Naively, it is possible to suggest that due to existence of the
second supersymmetry two first terms in Eq. (\ref{N=2_Action}) are
renormalized in the same way. As a consequence, the second
supersymmetry would lead to some restrictions to the
renormalization constants. However, this question is rather
subtle. Really, the regularized action proposed in this paper is
invariant under both supersymmetries of the ${\cal N}=2$
supersymmetric theory, but the gauge fixing term and the ghost
actions are invariant only under the manifest supersymmetry. It is
known \cite{Grigorian:1986yq} that any symmetry of the classical
action corresponds to a symmetry of the renormalized action and
the renormalized effective action. Nevertheless, we can not state
that in the case under consideration the invariance corresponding
to the second supersymmetry automatically fixes the
renormalization constant $Z_\Phi$. However, here we try to
understand what is needed for fixing the renormalization constant
$Z_\Phi$.

Using the background field method we rewrite the transformations
(\ref{Second_SUSY}) making the background--quantum splitting for
the gauge superfield and the chiral superfield $\Phi$:

\begin{eqnarray}
&&\qquad \delta (e^\Omega e^{\mbox{\scriptsize
\boldmath$\Omega$}})= i\eta^* e^\Omega e^{\mbox{\scriptsize
\boldmath$\Omega$}} (\Phi+\mbox{\boldmath$\Phi$});\qquad \delta
(e^{\mbox{\scriptsize \boldmath$\Omega$}^+} e^{\Omega^+}) = -i\eta
(\Phi^+ + \mbox{\boldmath$\Phi$}^+) e^{\mbox{\scriptsize
\boldmath$\Omega$}^+} e^{\Omega^+};\vphantom{\frac{1}{2}}\nonumber\\
&&\qquad\qquad\quad \delta(\Phi +\mbox{\boldmath$\Phi$}) =
-\frac{i}{2} \Big(\mbox{\boldmath$W$}^a + \frac{1}{8}
e^{-\mbox{\scriptsize \boldmath$\Omega$}}
\mbox{\boldmath$\bar\nabla$}^2 (e^{-2V} \mbox{\boldmath$\nabla$}^a
e^{2V}) e^{\mbox{\scriptsize \boldmath$\Omega$}} \Big) D_a\eta;\\
&& \delta\phi = -\frac{1}{4\sqrt{2}} \Big(\bar D^2 (\eta^*
e^{-2V_T} \widetilde\phi^*) -4m_0 \eta \phi\Big);\qquad
\delta\widetilde \phi = \frac{1}{4\sqrt{2}} \Big(\bar D^2 (\eta^*
e^{2V_T^t} \phi^*) -4m_0 \eta \widetilde\phi\Big),\qquad\nonumber
\end{eqnarray}

\noindent where $V_T$ is given by Eq.
(\ref{Total_Gauge_Superfield}) and

\begin{equation}
\mbox{\boldmath$W$}_a \equiv \frac{1}{8} \bar D^2
(e^{-2\mbox{\scriptsize \boldmath$V$}} D_a e^{2\mbox{\scriptsize
\boldmath$V$}}).
\end{equation}

\noindent These transformations can be obtained if we set

\begin{equation}\label{Second_SUSY_With_Background1}
\delta e^{\mbox{\scriptsize \boldmath$\Omega$}}= i\eta^*
e^{\mbox{\scriptsize \boldmath$\Omega$}}
\mbox{\boldmath$\Phi$};\qquad \delta e^{\mbox{\scriptsize
\boldmath$\Omega$}^+} = -i\eta \mbox{\boldmath$\Phi$}^+
e^{\mbox{\scriptsize \boldmath$\Omega$}^+};\qquad \delta
\mbox{\boldmath$\Phi$} = -\frac{i}{2} \mbox{\boldmath$W$}^a
D_a\eta
\end{equation}

\noindent for the background superfields and

\begin{eqnarray}\label{Second_SUSY_With_Background2}
&&\qquad\qquad\quad \delta e^\Omega=i\eta^* e^\Omega
e^{\mbox{\scriptsize \boldmath$\Omega$}} \Phi e^{-
\mbox{\scriptsize \boldmath$\Omega$}};\qquad\quad \delta
e^{\Omega^+}= -i\eta e^{- \mbox{\scriptsize \boldmath$\Omega$}^+}
\Phi e^{ \mbox{\scriptsize \boldmath$\Omega$}^+}
e^{\Omega^+};\vphantom{\frac{1}{2}}\nonumber\\
&& \qquad\qquad\qquad\qquad \delta\Phi = -\frac{i}{16}
e^{-\mbox{\scriptsize \boldmath$\Omega$}}
\mbox{\boldmath$\bar\nabla$}^2 (e^{-2V} \mbox{\boldmath$\nabla$}^a
e^{2V})
e^{\mbox{\scriptsize \boldmath$\Omega$}} D_a\eta;\nonumber\\
&& \delta\phi = -\frac{1}{4\sqrt{2}} \Big(\bar D^2 (\eta^*
e^{-2V_T} \widetilde\phi^*) -4m_0 \eta \phi\Big);\qquad
\delta\widetilde \phi = \frac{1}{4\sqrt{2}} \Big(\bar D^2 (\eta^*
e^{2V_T^t} \phi^*) -4m_0 \eta \widetilde\phi\Big)\qquad
\end{eqnarray}

\noindent for the quantum superfields.

We will show that if the effective action is invariant under the
hidden supersymmetry transformations
(\ref{Second_SUSY_With_Background1}), then

\begin{equation}\label{Z_Phi}
Z_\Phi=1.
\end{equation}

\noindent Actually, this is a manifestation of the general
statement that a symmetry can impose restrictions on the
renormalization constants. In this sense Eq. (\ref{Z_Phi}) is
similar, for example, to the equation $Z_3=1$ which follows from
the symmetry under transformations of the conformal group, see
e.g. \cite{Kataev:2013vua}. Note that the regularization proposed
in this paper is important, because it ensures the invariance of
the regularized action under the BRST and ${\cal N}=2$
supersymmetry transformations.

Now let us prove Eq. (\ref{Z_Phi}) assuming the invariance of the
effective action under the transformations
(\ref{Second_SUSY_With_Background1}). If the sources for the
hypermultiplet and quantum fields are set to 0, then the
invariance of the effective action under the transformations
(\ref{Second_SUSY_With_Background1}) can be expressed by the
equation

\begin{equation}\label{ST_Identity_Final}
0 = \mbox{tr}\int d^8x\, \Bigg\{\frac{\delta\Gamma}{\delta
\mbox{\boldmath$V$}}\, \delta_{\eta} \mbox{\boldmath$V$} +
\frac{\delta\Gamma}{\delta \mbox{\boldmath$\Phi$}}\cdot
\frac{D^2}{8\partial^2}\Big(-\frac{i}{2} \mbox{\boldmath$W$}^a
D_a\eta \Big) + \frac{\delta\Gamma}{\delta
\mbox{\boldmath$\Phi$}^+}\cdot \frac{\bar
D^2}{8\partial^2}\Big(\frac{i}{2} \mbox{\boldmath$\bar W$}^{\dot
a} \bar D_{\dot a}\eta^* \Big) \Bigg\},
\end{equation}

\noindent where $\delta_\eta \mbox{\boldmath$V$}$ is obtained from
the equation

\begin{equation}\label{Delta_V}
\delta_\eta (e^{2\mbox{\scriptsize \boldmath$V$}}) = i\eta^*
e^{2\mbox{\scriptsize \boldmath$V$}} \mbox{\boldmath$\Phi$} -i\eta
\mbox{\boldmath$\Phi$}^+ e^{2\mbox{\scriptsize \boldmath$V$}},
\end{equation}

\noindent which follows from Eq.
(\ref{Second_SUSY_With_Background1}). As a consequence,

\begin{equation}
\delta_\eta \mbox{\boldmath$V$} = \frac{i}{2} \eta^*
\mbox{\boldmath$\Phi$} - \frac{i}{2} \eta \mbox{\boldmath$\Phi$}^+
+ O(\mbox{\boldmath$V$}).
\end{equation}

\noindent Let us differentiate Eq. (\ref{ST_Identity_Final}) with
respect to $\mbox{\boldmath$V$}^B_y$ and
$\mbox{\boldmath$\Phi$}^{*A}_z$, where the subscripts denote
points in the superspace. Then, after setting all (background)
fields to 0 we obtain

\begin{equation}
\int d^8x\,\eta_x D^2\delta^8_{xz} \frac{\delta^2\Gamma}{\delta
\mbox{\boldmath$V$}^A_x \delta \mbox{\boldmath$V$}^B_y} = - \int
d^8x\, D^a\eta_x (D_a)_x \delta^8_{xy}
\frac{\delta^2\Gamma}{\delta \mbox{\boldmath$\Phi$}^{*A}_z \delta
\mbox{\boldmath$\Phi$}_{Bx}}.
\end{equation}

\noindent Because the effective action satisfies this relation, it
is possible to choose such a subtraction scheme in which this
relation is also valid for the renormalized action $S_R$:

\begin{equation}\label{Two_Point_Functions}
\int d^8x\,\eta_x D^2\delta^8_{xz} \frac{\delta^2 S_R}{\delta
\mbox{\boldmath$V$}^A_x \delta \mbox{\boldmath$V$}^B_y} = - \int
d^8x\, D^a\eta_x (D_a)_x \delta^8_{xy} \frac{\delta^2 S_R}{\delta
\mbox{\boldmath$\Phi$}^{*A}_z \delta \mbox{\boldmath$\Phi$}_{Bx}},
\end{equation}

\noindent where $S_R$ is constructed from the classical action by
substituting the bare fields by the renormalized ones. In
particular, a part of the renormalized action $S_R$ corresponding
to the two-point functions of the background fields has the form

\begin{eqnarray}\label{Renormalized_Two_Point_Functions}
&& \frac{1}{4} \int \frac{d^4p}{(2\pi)^4}\, d^4\theta\,\Big( -
\mbox{\boldmath$V$}_R^A(\theta,-p)
\partial^2 \Pi_{1/2} \mbox{\boldmath$V$}_R^A(\theta,p)
+ \mbox{\boldmath$\Phi$}_R^{*A}(\theta,-p)
\mbox{\boldmath$\Phi$}_R^A(\theta,p) \Big)\\
&& = \frac{1}{4} \int \frac{d^4p}{(2\pi)^4}\, d^4\theta\,\Big( -
\frac{1}{Z_3} \mbox{\boldmath$V$}^A(\theta,-p)
\partial^2 \Pi_{1/2} \mbox{\boldmath$V$}^A(\theta,p) +
(Z^{-2})_A{}^B \mbox{\boldmath$\Phi$}^{*A}(\theta,-p)
\mbox{\boldmath$\Phi$}_B(\theta,p) \Big),\qquad\nonumber
\end{eqnarray}

\noindent where $\partial^2\Pi_{1/2} \equiv - D^a \bar D^2 D_a/8$
is the supersymmetric projection operator and $(Z^{-1})_A{}^B$
denotes a matrix inverse to $Z_A{}^B$. Substituting this
expression into Eq. (\ref{Two_Point_Functions}) we obtain

\begin{equation}
\frac{1}{2 Z_3} \int d^8x\,\eta_x D^2\delta^8_{xz}
\partial^2\Pi_{1/2} \delta^8_{xy} \delta_A{}^B =
\frac{(Z^{-2})_A{}^B}{16} \int d^8x\, D^a\eta_x (D_a)_x
\delta^8_{xy} \bar D^2_x D^2_z \delta^8_{xz}.
\end{equation}

\noindent Integrating $\bar D_x^2$ by parts and taking into
account that the supersymmetry transformation parameter $\eta$ is
a space-time constant, after some simple transformations involving
the algebra of the covariant derivatives the integral in the right
hand side of this equation can be rewritten as

\begin{equation}
\int d^8x\, D^a\eta_x (\bar D^2 D_a)_x \delta^8_{xy} D^2_z
\delta^8_{xz} = 8 \int d^8x\, \eta_x (\partial^2 \Pi_{1/2})_x
\delta^8_{xy}  D^2_z \delta^8_{xz}.
\end{equation}

\noindent Therefore, the renormalization constants $Z_3$ and
$Z_A{}^B$ are related by the equation

\begin{equation}\label{Z_Relation}
Z_A{}^B = \sqrt{Z_3}\, \delta_A{}^B.
\end{equation}

\noindent Comparing this result with Eq. (\ref{ZAB}), we see that
the superfield $\mbox{\boldmath$\Phi$}$ is not renormalized,
$Z_{\Phi} = 1$, and we prove Eq. (\ref{Z_Phi}). As a result we get
the criterium whether the effective action is invariant under the
background hidden supersymmetry. This criterium is given by Eq.
(\ref{Z_Phi}). It means, in particular, that under this condition
the renormalization group function $\gamma_{\Phi}=0.$

\section{Finiteness of ${\cal N}=2$ supersymmetric theories beyond
the one-loop approximation and the NSVZ
$\beta$-function}\hspace{\parindent}
\label{Section_Finiteness_And_NSVZ}

Due to ${\cal N}=1$ supersymmetry a $\beta$-function of SYM
theories is related with the anomalous dimension of the matter
superfields by the NSVZ relation
\cite{Novikov:1983uc,Jones,Novikov:1985rd,Shifman:1986zi,Vainshtein:1986ja,Shifman:1985fi}.
For our purposes it is convenient to write it in the following
form \cite{Jack:1996vg}:

\begin{equation}\label{NSVZ_General_Beta}
\beta(\alpha_0) = - \frac{\alpha_0^2\Big(3 C_2 - T(R) + C(R)_i{}^j
\gamma_j{}^i(\alpha_0)/r \Big)}{2\pi(1- C_2\alpha_0/2\pi)},
\end{equation}

\noindent where

\begin{equation}
\alpha_0 = \frac{e_0^2}{4\pi}
\end{equation}

\noindent is a bare coupling constant, $\gamma_i{}^j(\alpha_0)$ is
the anomalous dimension of the matter superfields, and the
following notation is used:

\begin{eqnarray}\label{Notation}
&& \mbox{tr}\,(T^A T^B) \equiv T(R)\,\delta^{AB};\qquad
(T^A)_i{}^k
(T^A)_k{}^j \equiv C(R)_i{}^j;\nonumber\\
&& f^{ACD} f^{BCD} \equiv C_2 \delta^{AB};\qquad\quad r\equiv
\delta_{AA}.
\end{eqnarray}

\noindent The renormalization group functions in Eq.
(\ref{NSVZ_General_Beta}) are expressed in terms of the bare
coupling constant $\alpha_0$. These functions are defined
according to the following prescription:

\begin{equation}\label{RG_Functions_Bare}
\beta\Big(\alpha_0(\alpha,\Lambda/\mu)\Big) \equiv \frac{d
\alpha_0(\alpha,\Lambda/\mu)}{d\ln\Lambda}\Big|_{\alpha=\mbox{\scriptsize
const}};\qquad \gamma_i{}^j\Big(\alpha_0(\alpha,\Lambda/\mu)\Big)
\equiv -\frac{d \ln
Z_i{}^j(\alpha,\Lambda/\mu)}{d\ln\Lambda}\Big|_{\alpha=\mbox{\scriptsize
const}},
\end{equation}

\noindent where $\alpha$ denotes the renormalized coupling
constant $\alpha = e^2/4\pi$. The matter is that these are the
functions for that the NSVZ relation is obtained at least in the
Abelian case if the theory is regularized by higher derivatives
\cite{Stepanyantz:2011jy}.\footnote{For a fixed regularization the
renormalization group functions (\ref{RG_Functions_Bare}) are
scheme-independent, see e.g. \cite{Kataev:2013eta}.} Usually the
renormalization group functions are defined by a different way, in
terms of the renormalized coupling constant:

\begin{eqnarray}\label{RG_Functions_Renormalized}
&& \widetilde\beta\Big(\alpha(\alpha_0,\Lambda/\mu)\Big) \equiv
\frac{d\alpha(\alpha_0,\Lambda/\mu)}{d\ln\mu}\Big|_{\alpha_0=\mbox{\scriptsize
const}};\nonumber\\
&& \widetilde\gamma_i{}^j\Big(\alpha(\alpha_0,\Lambda/\mu)\Big)
\equiv \frac{d\ln Z_i{}^j(\alpha(\alpha_0,\Lambda/\mu),
\Lambda/\mu)}{d\ln\mu}\Big|_{\alpha_0=\mbox{\scriptsize const}}.
\end{eqnarray}

\noindent It is well-known that the $\beta$-function and the
anomalous dimension defined according to this prescription are
scheme-dependent. However \cite{Kataev:2013eta,Kataev:2013csa}, if
the boundary conditions

\begin{equation}\label{NSVZ_Scheme}
Z_3(\alpha,x_0) = 1;\qquad Z_i{}^j(\alpha,x_0)=1
\end{equation}

\noindent are imposed on the renormalization constants in an
arbitrary (but fixed) point $x_0
=\ln\Lambda/\mu_0$,\footnote{Although the first equation in Eq.
(\ref{NSVZ_Scheme}) looks similar to the condition $Z_3 =1$, which
can be imposed for obtaining the conformal symmetry limit of a
theory \cite{Kataev:2013vua}, there is a very important
difference: in Eq. (\ref{NSVZ_Scheme}) $Z_3=1$ only in a single
(but arbitrary) point $x_0$, while the conformal symmetry limit is
obtained if $Z_3=1$ for arbitrary values of $x$.} then the
renormalization group functions (\ref{RG_Functions_Renormalized})
coincide with the renormalization group functions
(\ref{RG_Functions_Bare}):

\begin{equation}
\widetilde\beta(\alpha) = \beta(\alpha);\qquad
\widetilde\gamma_i{}^j(\alpha) = \gamma_i{}^j(\alpha).
\end{equation}

\noindent This implies that the boundary conditions
(\ref{NSVZ_Scheme}) at least in the Abelian case give the NSVZ
scheme in all orders of the perturbation theory if the theory is
regularized by higher derivatives. This statement was verified by
the explicit three-loop calculations in Refs.
\cite{Kataev:2013eta,Kataev:2013csa}.

All features of the higher covariant derivative regularization, in
particular, factorization of integrals into integrals of (double)
total derivatives
\cite{Soloshenko:2003nc,Smilga:2004zr,Stepanyantz:2011jy}, which
gives the NSVZ relation for the renormalization group functions
(\ref{RG_Functions_Bare}) in the Abelian case, in the lowest loops
also take place in the non-Abelian case
\cite{Pimenov:2009hv,Stepanyantz:2011bz,Stepanyantz:2012zz,Stepanyantz:2012us}.
Although the all-loop derivation of the NSVZ relation is not so
far completed for SYM theories, it seems reasonable to suggest
that in the non-Abelian case the NSVZ relation is also obtained
for the renormalization group functions defined in terms of the
bare coupling constant with the higher covariant derivative
regularization. In this section we prove that under this
assumption the finiteness of ${\cal N}=2$ SYM theories beyond the
one-loop approximation can be very easily derived from the NSVZ
relation if the effective action is invariant under the
transformations (\ref{Second_SUSY_With_Background1}). Thus, the
regularization proposed in this paper possibly allows not only to
justify the non-renormalization theorems, but also to derive one
of them in the easiest way.

The main observation is that ${\cal N}=2$ SYM theories can be
considered as a special case of ${\cal N}=1$ supersymmetric
Yang--Mills theories, the representation $R$ for the matter
superfield being reducible and equal to the direct sum

\begin{equation}\label{R}
R = Adj + R_0 + \overline{R}_0.
\end{equation}

\noindent Here $Adj$ denotes the adjoint representation of the
gauge group corresponding to the superfield $\Phi$. (This
superfield together with the gauge superfield $V$ forms the ${\cal
N}=2$ gauge supermultiplet.) The fields in the representations
$R_0$ and $\overline{R}_0$ ($\phi$ and $\widetilde\phi$,
respectively) form the hypermultiplet.

Let us find the constants $C(R)_i{}^j$ and $T(R)$ for the
reducible representation (\ref{R}). For this purpose we note that
the generators of the considered representation can be written in
the form

\begin{equation}
T^A(R) = \left(
\begin{array}{ccc}
T^A(Adj) & 0 & 0\\
0 & T^A(R_0) & 0\\
0 & 0 & - (T^A(R_0))^t
\end{array}
\right).
\end{equation}

\noindent It is easy to see that for the adjoint representation
$T(Adj) = C_2$. Therefore,

\begin{equation}
T(R) = C_2 + 2 T(R_0).
\end{equation}

\noindent Also we obtain

\begin{equation}
C(R)_i{}^j = \left(
\begin{array}{ccc}
C_2 \cdot \delta_A^B & 0 & 0\\
0 & C(R_0) & 0\\
0 & 0 & C(R_0)
\end{array}
\right).
\end{equation}

\noindent We will prove that if the supersymmetric higher covariant
derivative regularization is used, the anomalous dimension of the
superfield $\Phi_A$ (defined in terms of the bare coupling constant)
is related with a $\beta$-function. This anomalous dimension is
calculated according to the following prescription:

\begin{equation}\label{Gamma_For_Phi}
\gamma(\alpha_0)_A{}^B \equiv -2\cdot \frac{d\ln
Z_A{}^B}{d\ln\Lambda}\Big|_{\alpha=\mbox{\scriptsize const}},
\end{equation}

\noindent where $\Phi_A = Z_A{}^B (\Phi_R)_B$ and the limit $m_0
\to 0$ is assumed. Then using Eq. (\ref{ZAB}) we obtain
\footnote{For the superfields $\Phi_A$ $Z$ in Eq.
(\ref{RG_Functions_Bare}) corresponds to $(Z^2)_A{}^B$. As a
consequence, we obtain the factor $2$ in Eq.
(\ref{Gamma_For_Phi}).}

\begin{eqnarray}
&& \gamma(\alpha_0)_A{}^B = -2\cdot \frac{d\ln
Z_A{}^B}{d\ln\Lambda} = -\frac{d\ln (Z_3
Z_\Phi)}{d\ln\Lambda}\delta_A{}^B = \Big(\frac{d\ln
\alpha_0/\alpha}{d\ln\Lambda} - \frac{d\ln Z_\Phi}{d
\ln\Lambda}\Big) \delta_A{}^B\nonumber\\
&&\qquad\qquad\qquad\qquad\qquad\qquad =
\Big(\alpha_0^{-1}\frac{d\alpha_0}{d\ln\Lambda} - \frac{d\ln
Z_\Phi}{d \ln\Lambda}\Big) \delta_A{}^B =
\Big(\frac{\beta(\alpha_0)}{\alpha_0} +
\gamma_\Phi(\alpha_0)\Big)\delta_A{}^B,\qquad\quad
\end{eqnarray}

\noindent where

\begin{equation}
\gamma_\Phi\Big(\alpha_0(\alpha,\Lambda/\mu)\Big) \equiv -\frac{d
\ln Z_\Phi(\alpha,\Lambda/\mu)}{d\ln\Lambda}
\Big|_{\alpha=\mbox{\scriptsize const}}
\end{equation}

\noindent is the anomalous dimension of the superfield $\Phi$
defined in terms of the bare coupling constant.

The anomalous dimension of the hypermultplet can be expressed
through $\gamma_\Phi$ using Eqs. (\ref{Z_Hypermultiplet}):

\begin{equation}\label{Hypermultiplet_Renormalization}
(\gamma_\phi)_i{}^j \equiv - \frac{d\ln Z_\phi}{d\ln\Lambda} \cdot
\delta_i^j = \frac{1}{2}\, \frac{d\ln Z_\Phi}{d\ln\Lambda} \cdot
\delta_i^j = - \frac{1}{2}\,\gamma_\Phi(\alpha_0)\cdot \delta_i^j.
\end{equation}

\noindent Therefore, the anomalous dimension can be written as

\begin{equation}\label{Exact_Gamma}
\gamma_i{}^j(\alpha_0) = \left(
\begin{array}{ccc}
\left(\beta(\alpha_0)/\alpha_0 + \gamma_\Phi(\alpha_0)\right) \cdot \delta_A^B & 0 & 0\\
0 & -\gamma_\Phi(\alpha_0)/2\cdot\delta_i^j & 0\\
0 & 0 & -\gamma_\Phi(\alpha_0)/2\cdot\delta_i^j
\end{array}
\right),
\end{equation}

Substituting the expressions for $T(R)$, $C(R)_i{}^j$, and
$\gamma_i{}^j$ in Eq. (\ref{NSVZ_General_Beta}) we obtain

\begin{equation}
\beta(\alpha_0) = - \frac{\alpha_0^2\Big(2 C_2 - 2T(R_0) + C_2
\Big(\beta(\alpha_0)/\alpha_0 + \gamma_\Phi(\alpha_0)\Big) -
T(R_0)\gamma_\Phi(\alpha_0)\Big)}{2\pi(1- C_2\alpha_0/2\pi)}.
\end{equation}

\noindent Solving this equation for $\beta(\alpha_0)$, after some
simple transformations we find that a $\beta$-function of the
considered theory is

\begin{equation}\label{Exact_Beta_For_N=2}
\beta(\alpha_0) = - \frac{\alpha_0^2}{\pi}\Big(C_2 -
T(R_0)\Big)\Big(1 + \frac{1}{2}\,\gamma_\Phi(\alpha_0)\Big).
\end{equation}

Thus, we see that in the theory under consideration the higher
loop structure of the NSVZ $\beta$-function is determined by the
function $\gamma_{\Phi}(\alpha_0)$. If $\gamma_\Phi=0$, the
expression (\ref{Exact_Beta_For_N=2}) contains only $\alpha_0^2$.
Therefore, the NSVZ $\beta$-function for an arbitrary
renormalizable ${\cal N}=2$ supersymmetric Yang--Mills theory does
not vanish only in the one-loop approximation and coincides with
conventional one-loop $\beta$-function:

\begin{equation}\label{Exact_One-Loop_Beta_For_N=2}
\beta(\alpha_0) = - \frac{\alpha_0^2}{\pi}\Big(C_2 - T(R_0)\Big).
\end{equation}

\noindent As the result, we can conclude that the ${\cal N}=2$
non-renormalization theorem is equivalent to the statement
$\gamma_{\Phi}=0$.

The equality $\gamma_\Phi=0$ follows from the invariance of the
renormalized action under the background transformations of the
hidden supersymmetry (see Section 4). It is evident that this
invariance takes place if both a regularization and a gauge fixing
procedure are invariant under the complete ${\cal N}=2$
supersymmetry. However, the considered gauge fixing term is
invariant only under the manifest supersymmetry. Nevertheless, we
can present here some indirect arguments in favor of this
equality. It is known that the $\beta$-function is gauge
independent if the minimal substraction scheme is used for
renormalization (see e.g. \cite{Voronov:1982wv}). Therefore, if
there exists a completely ${\cal N}=2$ supersymmetric gauge, then
the regularized effective action will be invariant under the same
amount of supersymmetries as the classical action and according to
(\ref{Z_Phi}) one gets $\gamma_\Phi=0$ and the $\beta$-function
vanishes beyond one-loop. Since the $\beta$-function is gauge
invariant, the same result will be valid in any gauge, in
particular in the gauge used in this paper. But the completely
${\cal N}=2$ invariant gauge does actually exist
\cite{Howe:1983sr}. However, the ${\cal N}=2$ supersymmetric gauge
used for derivation of Eq. (\ref{Z_Phi}) was formulated in terms
of ${\cal N}=2$ superfields while the proposed regularization is
formulated in terms of ${\cal N}=1$ superfields and it is unclear
whether these gauge and regularization are consistent one with
another.

Thus, if we accept that $\gamma_\Phi=0$, then up to a possibility
of making finite renormalizations we obtain the following values
of the renormalization constants (exactly in all orders):

\begin{equation}\label{Renormalization_Constants}
Z_{3} = 1+ \frac{\alpha}{\pi}\Big(C_2 -
T(R_0)\Big)\ln\frac{\Lambda}{\mu};\quad\ Z_\Phi=1;\quad\ Z_\phi =
1;\quad\ Z_m =1,
\end{equation}

\noindent where $\mu$ is a renormalization parameter. Values of
$Z_V$, $Z_c$ and $Z_{\bar c}$ are not so far defined. In the next
section we describe the one-loop calculation, which allows to find
values of these renormalization constants in the one-loop
approximation.

\section{One-loop renormalization with the higher covariant
derivative regularization}
\hspace{\parindent}\label{Section_One-Loop}

In this section we calculate one-loop divergences using the
version of the higher covariant derivative regularization
constructed in this paper.

\begin{figure}[h]
\begin{picture}(0,2)
\put(4,0){\includegraphics[scale=0.4]{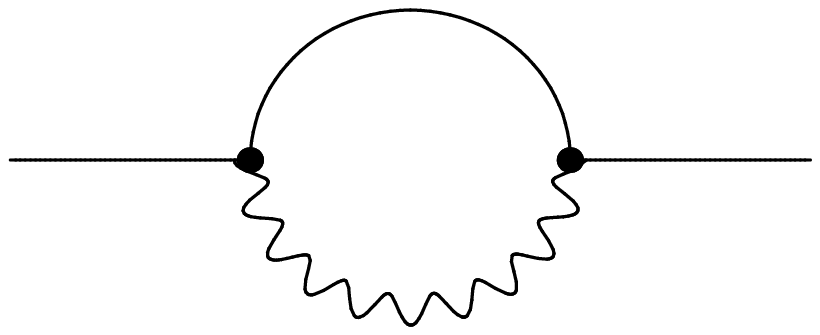}}
\put(6.3,-0.1){$V$} \put(3.5,0.9){$\phi$\ or\ $\widetilde\phi$}
\put(8,0){\includegraphics[scale=0.4]{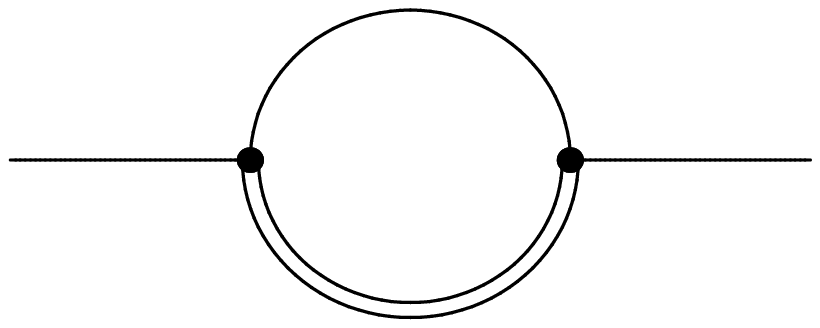}}
\put(10.3,-0.1){$\Phi$}
\end{picture}
\vspace*{1mm} \caption{Diagrams giving a two-point Green function
of the hypermultiplet in the one-loop
approximation.}\label{Figure_Hypermultiplet Renormalization}
\end{figure}

In the one-loop approximation the point Green function of the
hypermultiplet, $G_i{}^j$, defined by the equation

\begin{equation}
\Gamma^{(2)}_\phi = \frac{1}{4} \int \frac{d^4p}{(2\pi)^4}\,
d^4\theta\,\Big(\phi^*(\theta,-p)^i \phi(\theta,p)_j + \widetilde
\phi^*(\theta,-p)^i \widetilde \phi(\theta,p)_j\Big)
G_i{}^j(\alpha_0,\Lambda/p,m_0/\Lambda),
\end{equation}

\noindent can be obtained by calculating two diagrams presented in
Fig. \ref{Figure_Hypermultiplet Renormalization}. It is easy to
see that these diagrams cancel each other:

\begin{eqnarray}
&& G_i{}^j(\alpha_0,\Lambda/p,m_0/\Lambda) = \delta_i^j -
C(R)_i{}^j \int \frac{d^4q}{(2\pi)^4}
\frac{2e_0^2}{q^2(1+q^2/\Lambda^2)\Big((q+p)^2+m_0^2\Big)}\nonumber\\
&& + C(R)_i{}^j \int \frac{d^4q}{(2\pi)^4}
\frac{2e_0^2}{q^2(1+q^2/\Lambda^2)\Big((q+p)^2+m_0^2\Big)} +
O(e_0^4) =\delta_i^j + O(e_0^4)
\end{eqnarray}

\noindent This result is in a complete agreement with Eq.
(\ref{Renormalization_Constants}). (For finite terms we always
choose the renormalization constants equal to 1.) Using exactly
the same arguments we prove that the Pauli--Villars fields
$\varphi$, $\widetilde\varphi$, $\phi_I$, and $\widetilde\phi_I$
are not renormalized in the one-loop approximation.

\begin{figure}[h]
\begin{picture}(0,2.0)
\put(1,0){\includegraphics[scale=0.4]{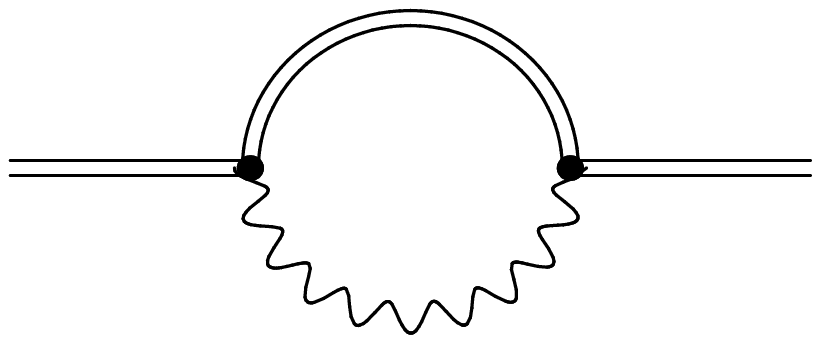}}
\put(1,0.9){$\Phi$} \put(3.2,-0.2){$V$}
\put(5,0.05){\includegraphics[scale=0.4]{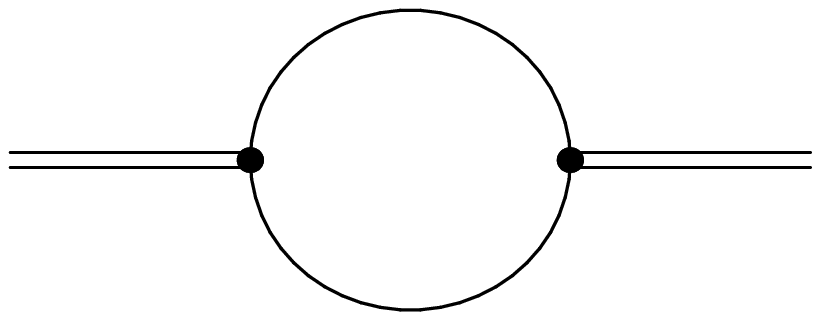}}
\put(5.4,1.6){$\phi,\widetilde\phi,\phi_I,\widetilde\phi_I,\varphi,\widetilde\varphi$}
\put(9.4,0){\includegraphics[scale=0.4]{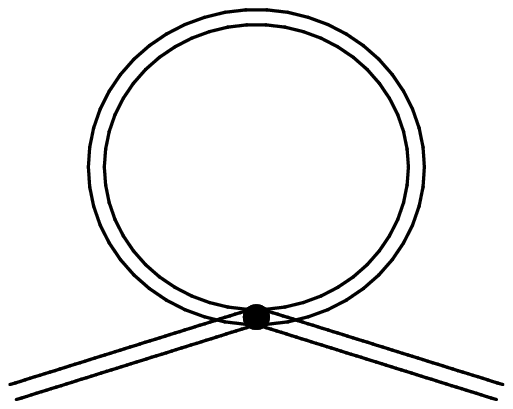}}
\put(9.6,1.6){$\Phi$}
\put(12.4,0){\includegraphics[scale=0.4]{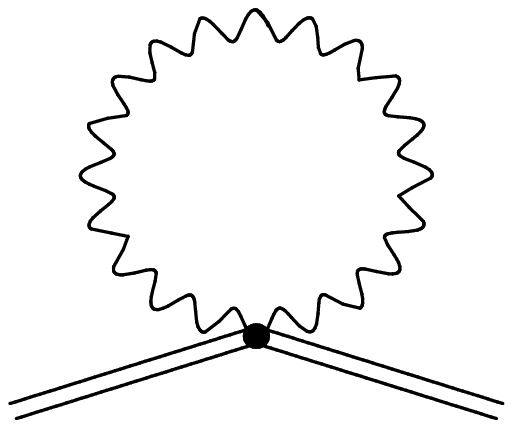}}
\put(12.6,1.6){$V$}
\end{picture}
\vspace*{1mm} \caption{Diagrams giving a two-point Green function
of the superfield $\Phi$ in the one-loop
approximation.}\label{Figure_Phi Renormalization}
\end{figure}

The two-point Green function of the superfield $\Phi$ in the
one-loop approximation is determined by diagrams presented in Fig.
\ref{Figure_Phi Renormalization}. Calculating these diagrams we
obtain the function $G$ defined by the equation

\begin{equation}
\Gamma^{(2)}_\Phi = \frac{1}{2e_0^2} \mbox{tr} \int
\frac{d^4p}{(2\pi)^4}\, d^4\theta\,\Phi^+(\theta,-p)
\Phi(\theta,p) G(\alpha_0,\Lambda/p,m_0/\Lambda),
\end{equation}

\noindent where $\Phi = e_0 \Phi_A t^A$ and, for simplicity, we
assume that the gauge group is simple. (The background superfield
$\mbox{\boldmath$\Phi$}$ is omitted, because the effective action
depends only on $\Phi+\mbox{\boldmath$\Phi$}$.) We are interested
in the divergent part of the function $G$. Taking into account
that in the one-loop approximation logarithmically divergent terms
are proportional to $\ln\Lambda$, it can be found by
differentiating the result for the function $\ln G$ (in the
one-loop approximation this is equivalent to differentiating the
function $G$) with respect to $\ln\Lambda$ in the limit of the
vanishing external momentum:

\begin{eqnarray}\label{G_With_P=0}
&& \frac{d\ln G}{d\ln\Lambda}\Big|_{p\to 0} =
\frac{d}{d\ln\Lambda} \Bigg\{ e_0^2 \int
\frac{d^4k}{(2\pi)^4}\Bigg( 2 T(R_0)\Big(\frac{1}{(k^2 + m_0^2)^2}
-
\sum\limits_{I=1}^n c_I \frac{1}{(k^2 + M_I^2)^2} \Big)\nonumber\\
&& - 2C_2 \Big(\frac{1}{k^4} - \frac{1}{(k^2+M_0^2)^2}\Big) +
\frac{2 C_2}{\Lambda^2 k^2 (1+k^2/\Lambda^2)} - \frac{2
C_2}{\Lambda^2 k^2 (1+k^2/\Lambda^2)}\Bigg) +
O(e_0^4)\Bigg\}.\qquad
\end{eqnarray}

\noindent Here the first term (proportional to $T(R_0)$) is a
contribution of the hypermultiplet and the corresponding
Pauli--Villars fields (the second diagram in Fig. \ref{Figure_Phi
Renormalization}). The second term consists of the contributions
of the first diagram in Fig. \ref{Figure_Phi Renormalization} and
the Pauli--Villars fields $\varphi$ and $\widetilde\varphi$ (the
second diagram in Fig. \ref{Figure_Phi Renormalization}). The
third and the fourth terms correspond to the third and the fourth
diagrams in Fig. \ref{Figure_Phi Renormalization}, respectively,
and cancel each other. Thus, we see that this expression is finite
and can be easily calculated.\footnote{The considered Green
function is also finite in the infrared limit if $p\ne 0$. In Eq.
(\ref{G_With_P=0}) it is possible to take the limit $p\to 0$ due
to the derivative with respect to $\ln\Lambda$.} In order to do
this we note that Eq. (\ref{G_With_P=0}) can be rewritten as an
integral over a double total derivative:

\begin{eqnarray}\label{G_Derivative}
&& \frac{d\ln G}{d\ln\Lambda}\Big|_{p\to 0} = - e_0^2 \int
\frac{d^4k}{(2\pi)^4}\frac{d}{d\ln\Lambda}
\frac{\partial}{\partial k^\mu} \frac{\partial}{\partial k_\mu}
\Bigg( \frac{T(R_0)}{2k^2}
\Big(\ln(k^2+m_0^2) -\sum\limits_{I=1}^n c_I \ln(k^2 + M_I^2)\Big) \nonumber\\
&& - \frac{C_2}{2k^2} \Big(\ln k^2 -\ln (k^2+M_0^2)\Big) \Bigg) +
O(e_0^4).
\end{eqnarray}

\noindent Taking into account that the Pauli--Villars masses $M_I$
and $M_0$ are proportional to the parameter $\Lambda$, we easily
obtain (setting $m_0=0$)

\begin{equation}\label{Phi_Anomalous_Dimension}
\gamma(\alpha_0) = \frac{d\ln G}{d\ln\Lambda}\Big|_{p\to 0;\
m_0=0} = \frac{e_0^2}{4\pi^2} (T(R_0)-C_2) + O(e_0^4) =
\frac{\alpha_0}{\pi}(T(R_0)-C_2) + O(\alpha_0^2).
\end{equation}

\noindent As a consequence,

\begin{equation}
G = 1 + \frac{\alpha_0}{\pi} (T(R_0)-C_2) \ln\Lambda +
\mbox{finite terms} + O(\alpha_0^2).
\end{equation}

\begin{figure}[h]
\begin{picture}(0,2)
\put(6.3,0.2){\includegraphics[scale=0.4]{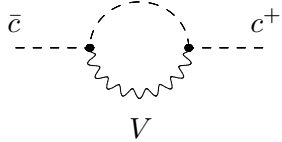}}
\put(6.2,1.1){$\bar c$} \put(7.8,-0.3){$V$} \put(9.4,1.1){$c^+$}
\end{picture}
\vspace*{1mm} \caption{This diagram gives a two-point Green
function of the Faddeev--Popov ghost superfields in the one-loop
approximation.}\label{Figure_Ghost_Renormalization}
\end{figure}

Evidently, the Nielsen--Kallosh ghosts are not renormalized,
because they interact only with the background gauge superfield.
The one-loop renormalization of the Faddeev--Popov ghosts can be
found by calculating a diagram presented in Fig.
\ref{Figure_Ghost_Renormalization}.

\noindent It is easy to see that contributions of the various
ghosts fields cancel each other and this diagram is convergent and
gives the vanishing contribution. Therefore, in the one-loop
approximation it is possible to choose $Z_c=1$.

\begin{figure}[h]
\begin{picture}(0,4.5)
\put(0.3,2.7){\includegraphics[scale=0.4]{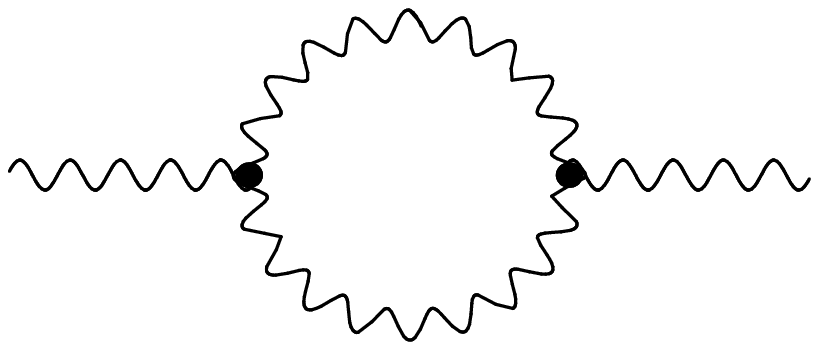}}
\put(-0.3,3.6){\mbox{\boldmath$V$}\ or\ $V$} \put(1.8,2.1){$V$}
\put(4.3,2.7){\includegraphics[scale=0.4]{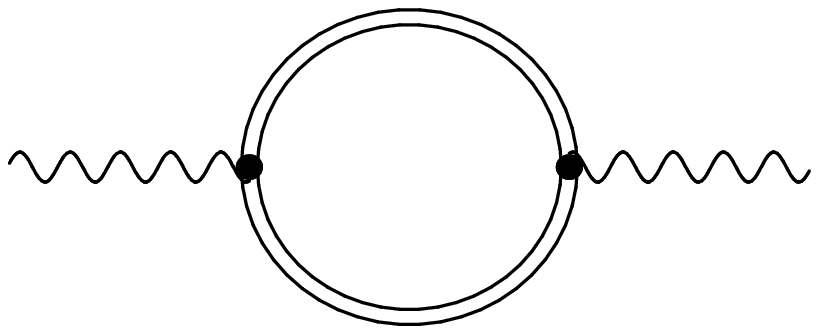}}
\put(5.9,2.1){$\Phi$}
\put(8.3,2.7){\includegraphics[scale=0.4]{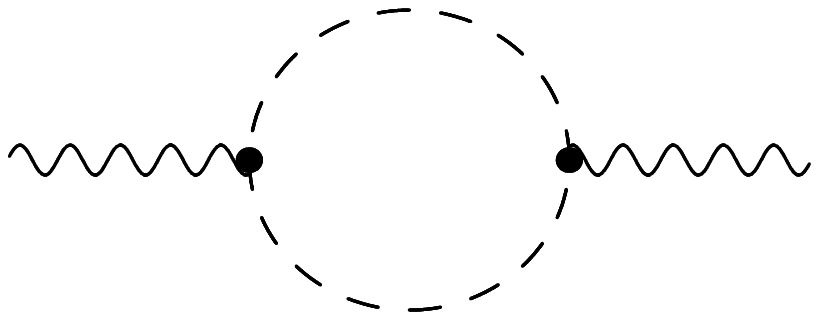}}
\put(9.5,2.1){$c,\bar c,C$}
\put(12.3,2.7){\includegraphics[scale=0.4]{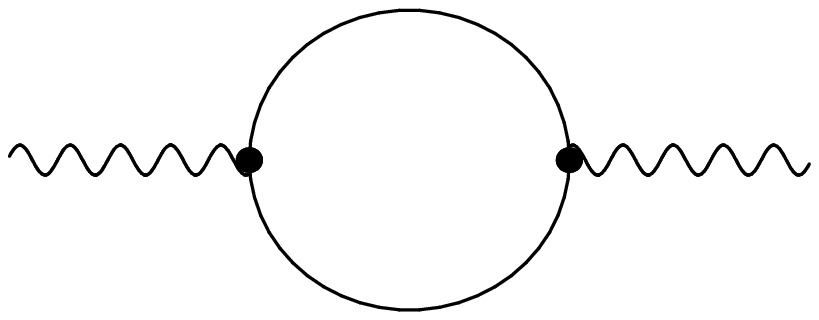}}
\put(12.8,2.1){$\phi,\widetilde\phi,\phi_I,\widetilde\phi_I,\varphi,\widetilde\varphi$}
\put(0.7,0){\includegraphics[scale=0.4]{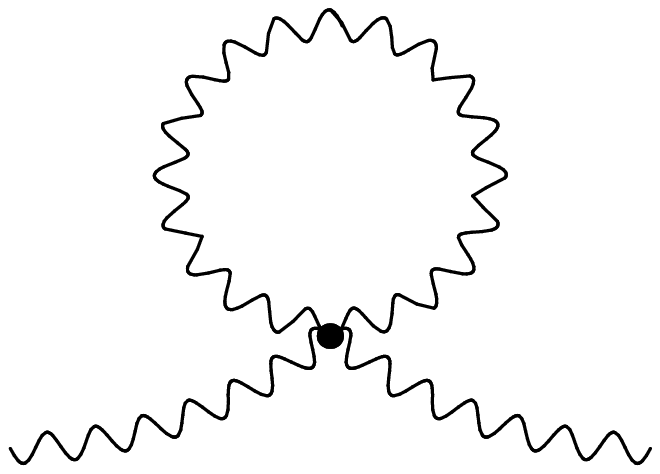}}
\put(0.0,0.3){\mbox{\boldmath$V$}\ or\ $V$}
\put(4.7,0){\includegraphics[scale=0.4]{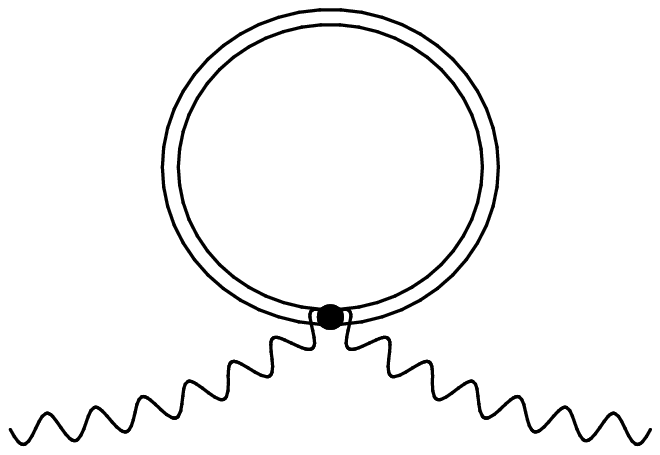}}
\put(8.7,0){\includegraphics[scale=0.4]{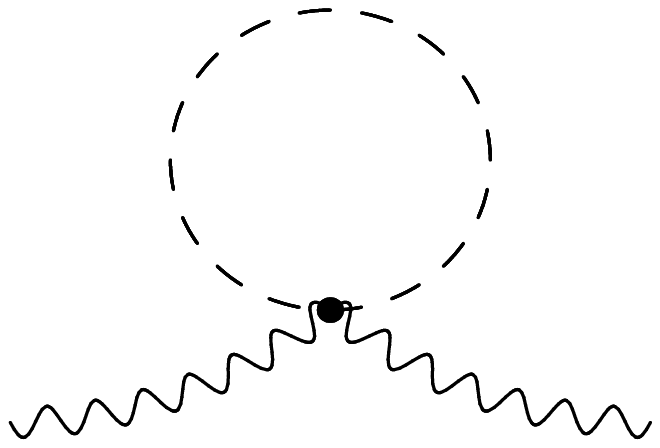}}
\put(12.7,0){\includegraphics[scale=0.4]{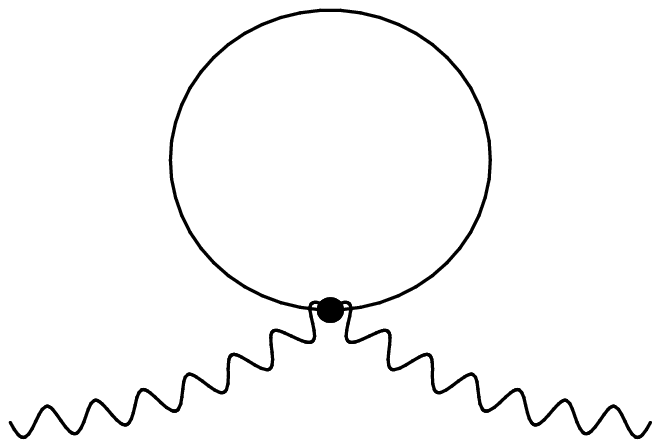}}
\end{picture}
\vspace*{1mm} \caption{Diagrams giving a two-point Green function
of the background gauge superfield $\mbox{\boldmath$V$}$ in the
one-loop approximation. These diagrams without the ones with a
loop of the Nielsen--Kallosh ghosts $C$ also give a one-loop
renormalization of the quantum gauge
field.}\label{Figure_Alpha_Renormalization}
\end{figure}

Renormalization of the coupling constant can be investigated by
calculating the two-point Green function of the background gauge
superfield. Due to the Slavnov--Taylor identity
\cite{Taylor:1971ff,Slavnov:1972fg} this Green function is
transversal:

\begin{equation}\label{D_Definition}
\Gamma^{(2)}_{\mbox{\scriptsize \boldmath$V$}} = - \frac{1}{8\pi}
\mbox{tr}\int \frac{d^4p}{(2\pi)^4}\,d^4\theta\,
\mbox{\boldmath$V$}(\theta,-p)\,\partial^2\Pi_{1/2}
\mbox{\boldmath$V$}(\theta,p)\,
d^{-1}(\alpha_0,\Lambda/p,m_0/\Lambda).
\end{equation}

\noindent In the one-loop approximation the function $d^{-1}$ can
be obtained by calculating the diagrams presented in Fig.
\ref{Figure_Alpha_Renormalization}. The result has the following
form:

\begin{eqnarray}\label{D-1_Derivative}
&& \frac{d}{d\ln\Lambda}(d^{-1}-\alpha_0^{-1})\Big|_{p\to 0} =
4\pi\cdot \frac{d}{d\ln\Lambda} \int \frac{d^4k}{(2\pi)^4}\Bigg(
C_2 \Big(\frac{1}{k^4} +
\frac{2}{\Lambda^4 (1+k^2/\Lambda^2)^2}\Big)\nonumber\\
&& - C_2 \Big(\frac{1}{k^4} + \frac{2}{\Lambda^4
(1+k^2/\Lambda^2)^2}\Big) +2 T(R_0)\Big(\frac{1}{(k^2 + m_0^2)^2}
-\sum\limits_{I=1}^n c_I \frac{1}{(k^2 + M_I^2)^2} \Big)\nonumber\\
&& - 2C_2 \Big(\frac{1}{k^4} - \frac{1}{(k^2+M_0^2)^2}\Big) \Bigg)
+ O(e_0^2).
\end{eqnarray}

\noindent The diagrams containing an internal loop of the quantum
gauge superfield $V$ (the first column in Fig.
\ref{Figure_Alpha_Renormalization}) give a vanishing contribution
in the limit $p\to 0$. The diagrams with an internal loop of
$\Phi$ (the second column in Fig.
\ref{Figure_Alpha_Renormalization}) give the first term in Eq.
(\ref{D-1_Derivative}). This term is exactly canceled by a
contribution of the diagrams containing a loop of the
Nielsen--Kallosh ghosts $C$, which is given by the second term in
Eq. (\ref{D-1_Derivative}). The third term in Eq.
(\ref{D-1_Derivative}) corresponds to the contribution of the
hypermultiplet $\phi,\widetilde\phi$ and its Pauli--Villars fields
$\phi_I,\widetilde\phi_I$ (the fourth column in Fig.
\ref{Figure_Alpha_Renormalization}). The last term in Eq.
(\ref{D-1_Derivative}) consists of the Faddeev--Popov ghosts
($c,\bar c$) contribution and the contribution of the
Pauli--Villars fields $\varphi,\widetilde\varphi$.

Taking into account that

\begin{equation}
\frac{d}{d\ln\Lambda}(d^{-1}-\alpha_0^{-1})\Big|_{p\to 0;\ m_0=0}
= - \frac{d}{d\ln\Lambda} (\alpha_0^{-1})\Big|_{m_0=0} =
\frac{\beta(\alpha_0)}{\alpha_0^2},
\end{equation}

\noindent we obtain that a $\beta$-function of the considered
theory is given by integrals of double total derivatives:

\begin{eqnarray}
&& \frac{\beta(\alpha_0)}{\alpha_0^2} = - 4\pi \int
\frac{d^4k}{(2\pi)^4}\frac{d}{d\ln\Lambda}
\frac{\partial}{\partial k^\mu} \frac{\partial}{\partial k_\mu}
\Bigg( \frac{T(R_0)}{2k^2}
\Big(\ln(k^2+m_0^2) -\sum\limits_{I=1}^n c_I \ln(k^2 + M_I^2)\Big) \nonumber\\
&& - \frac{C_2}{2k^2} \Big(\ln k^2 -\ln (k^2+M_0^2)\Big)
\Bigg)\Big|_{m_0=0}  + O(\alpha_0)= \frac{1}{\pi}\Big(T(R_0) -
C_2\Big) + O(\alpha_0).
\end{eqnarray}

\noindent In the considered approximation this result agrees with
the exact expression (\ref{Exact_Beta_For_N=2}). Comparing it with
Eq. (\ref{Phi_Anomalous_Dimension}) we verify Eq.
(\ref{Exact_Gamma}) in the considered (one-loop) approximation.

Due to the Slavnov--Taylor identity the two-point Green function
of the quantum gauge superfield is also transversal:

\begin{equation}\label{D_Quantum}
\Gamma^{(2)}_V - S^{(2)}_{\mbox{\scriptsize gf}}= - \frac{1}{8\pi}
\mbox{tr}\int \frac{d^4p}{(2\pi)^4}\,d^4\theta\,
V(\theta,-p)\,\partial^2\Pi_{1/2} V(\theta,p)\,
d_{\mbox{\scriptsize q}}^{-1}(\alpha_0,\Lambda/p,m_0/\Lambda).
\end{equation}

\noindent The function $d_q^{-1}$ can be also found by calculating
the diagrams presented in Fig. \ref{Figure_Alpha_Renormalization}.
The only difference is that the Nielsen--Kallosh ghosts $C$ do no
contribute to the renormalization of the quantum gauge superfield.
The result has the following form:

\begin{eqnarray}\label{DQ-1_Derivative}
&& \frac{d}{d\ln\Lambda}(d_{\mbox{\scriptsize
q}}^{-1}-\alpha_0^{-1})\Big|_{p\to 0} = 4\pi\cdot
\frac{d}{d\ln\Lambda} \int \frac{d^4k}{(2\pi)^4}\Bigg( C_2
\Big(\frac{1}{k^4} +
\frac{2}{\Lambda^4 (1+k^2/\Lambda^2)^2}\Big)\nonumber\\
&& - C_2 \Big(\frac{3}{k^4} + \frac{2}{\Lambda^4
(1+k^2/\Lambda^2)^2}\Big) +2 T(R_0)\Big(\frac{1}{(k^2 + m_0^2)^2}
- \sum\limits_{I=1}^n c_I \frac{1}{(k^2 + M_I^2)^2} \Big)
\nonumber\\
&& +  \frac{2C_2}{(k^2+M_0^2)^2} \Bigg) + O(e_0^2).
\end{eqnarray}

\noindent The expression in the right hand side of this equation
is finite at finite values of $\Lambda$ and coincides with the
corresponding expression in Eq. (\ref{D-1_Derivative}). The
contributions of the superfield $\Phi$, the hypermultiplet (with
the corresponding Pauli--Villars fields), and the Pauli--Villars
fields $\varphi$ and $\widetilde\varphi$ are calculated exactly as
earlier. However, contributions of the quantum gauge superfield
and ghosts are different, if the external lines correspond to the
quantum gauge superfield $V$. As we have already mentioned above,
the Nielsen--Kallosh ghosts do not contribute to the
renormalization of the quantum gauge superfield, because their
action depends only on the background gauge superfield. The
Faddeev--Popov ghosts give only noninvariant terms proportional to
$\mbox{tr}\, V^2$, which exactly cancel similar terms coming from
the diagrams with a loop of the quantum gauge superfield. The
diagrams with a loop of the quantum gauge superfield also give
invariant contribution, which is given by the second term in Eq.
(\ref{DQ-1_Derivative}).

It is also expedient to compare Eqs. (\ref{G_Derivative}),
(\ref{D-1_Derivative}), and (\ref{DQ-1_Derivative}). For this
purpose we write the one-loop divergences of the considered two
point functions in the following form (taking into account that
the effective action depends on the superfield $\Phi_T = \Phi +
\mbox{\boldmath$\Phi$}$):

\begin{eqnarray}\label{Divergences}
&&\hspace*{-6mm} \frac{1}{2 e_0^2} \mbox{tr} \int d^4x\,
d^4\theta\,\Big(- \mbox{\boldmath$V$} \partial^2 \Pi_{1/2}
\mbox{\boldmath$V$} - V
\partial^2 \Pi_{1/2} V + \Phi_T^+ \Phi_T \Big) \ln\Lambda \Bigg\{-
e_0^2 \int \frac{d^4k}{(2\pi)^4}\frac{d}{d\ln\Lambda}
\frac{\partial}{\partial k^\mu} \frac{\partial}{\partial k_\mu}
\nonumber\\
&&\hspace*{-6mm} \times \Bigg( \frac{T(R_0)}{2k^2}
\Big(\ln(k^2+m_0^2) -\sum\limits_{I=1}^n c_I \ln(k^2 + M_I^2)\Big)
- \frac{C_2}{2k^2} \Big(\ln k^2 -\ln (k^2+M_0^2)\Big) \Bigg) +
O(e_0^4)\Bigg\}.\qquad
\end{eqnarray}

\noindent From this equation we see that the regularization
constructed in this paper in the considered approximation allows
to obtain the manifestly ${\cal N}=2$ supersymmetric effective
action, although the gauge fixing procedure is not ${\cal N}=2$
supersymmetric. From Eq. (\ref{Divergences}) we also conclude that
the superfield $\Phi$ is not renormalized, $Z_\Phi=1$, because all
divergences are absorbed into the coupling constant
renormalization. This completely agrees with Eq. (\ref{Z_Phi}).
Certainly, it is possible to make a finite renormalization of the
superfield $\Phi$. However, such a finite renormalization destroys
${\cal N}=2$ supersymmetry and we will not make it. Moreover, we
see that the quantum field $V$ is not renormalized in the one-loop
approximation, so that it is possible to choose $Z_V=1$.

Thus, we have verified that the proposed regularization does
regularize the one-loop divergences and gives the correct values
of the renormalization group functions in the one-loop
approximation. In particular, we confirm Eq.
(\ref{Renormalization_Constants}) by the explicit calculation in
the one-loop approximation and also obtain

\begin{equation}
Z_V =1;\qquad Z_c Z_{\bar c}=1.
\end{equation}

\section{Conclusion}
\hspace{\parindent}

We have proposed a new version of the higher covariant derivative
regularization for general ${\cal N}=2$ SYM theories formulated in
terms of ${\cal N}=1$ superfields. At the classical level such
theories are manifestly invariant under ${\cal N}=1$ supersymmetry
by construction, but these theories are also invariant under
additional hidden on-shell supersymmetry.

For calculation of quantum corrections it is convenient to define
the effective action in the framework of the background field
method and fix a gauge without breaking the background gauge
invariance. In order to regularize the theory by higher covariant
derivatives, we constructed the gauge invariant higher derivative
functional which is invariant under the same amount of
supersymmetries as the classical action. Adding this functional to
the classical action we regularize all divergences beyond the
one-loop approximation in the gauge invariant and ${\cal N}=2$
supersymmetric way. The remaining one-loop divergences are
regularized by inserting appropriate Pauli--Villars determinants
into the generating functional. We show that these determinants
preserve all supersymmetries of the classical action by
construction. As a result, the hidden supersymmetry is broken only
by the gauge fixing procedure. In this paper we have found that if
the effective action is invariant under the background
transformation of the hidden supersymmetry, the renormalization of
the coupling constant is related with the renormalization of the
superfields $\Phi_A$ (or, equivalently, the superfield $\Phi = e_0
\Phi_A t^A$ is unrenormalized, $Z_\Phi=1$). The exact NSVZ
$\beta$-function is naturally obtained with help of the higher
derivative regularization. Thus, it is possible to use the
relation (\ref{Exact_Beta_For_N=2}), which follows from the exact
NSVZ $\beta$-function. This, in turn, implies that the higher loop
structure of exact NSVZ $\beta$-function is determined by the
anomalous dimension $\gamma_{\Phi}(\alpha_0)$. If the function
$\gamma_{\Phi}$ vanishes, the NSVZ $\beta$-function is reduced to
a purely one-loop expression. Therefore, the equality
$\gamma_{\Phi}=0$ discussed above can be considered as the exact
criterium of finiteness of ${\cal N}=2$ SYM theories with matter
beyond the one-loop approximation. We want to emphasize once more,
that all previous proofs of the ${\cal N}=2$ non-renormalization
theorem were based on the assumption of existence of a
regularization preserving all symmetries of the classical action
in an arbitrary loop. However, all known regularizations do not
satisfy this assumption. In this paper we actually presented such
a regularization and showed how it works.

Also, we would like to point out that a completely off-shell
${\cal N}=2$ supersymmetric regularization can in principle be
developed within the harmonic superfield approach to ${\cal N}=2$
supersymmetric theories \cite{Galperin:2001}. This approach allows
to formulate ${\cal N}=2$ SYM theories in terms of off-shell
${\cal N}=2$ superfields. Moreover, the background field formalism
and off-shell ${\cal N}=2$ supersymmetric gauge fixing procedure
are developed in the harmonic superfield approach
\cite{Buchbinder:1997ya,Buchbinder:1997ib}. Therefore, for
constructing a manifestly ${\cal N}=2$ supersymmetric
regularization it is necessary to construct an appropriate gauge
invariant higher derivative functional in terms of harmonic
superfields. We plan to study this problem in a forthcoming work.

\bigskip
\bigskip

\noindent {\Large\bf Acknowledgements.}

\bigskip

\noindent The authors are very grateful to A.L. Kataev,
A.A.Slavnov, and I.V.Tyutin for valuable discussions. The work of
I.L.B was partially supported by RFBR grant, project No
12-02-00121, RFBR-DFG grant, project No 13-02-91330, grant for
LRSS, project No 88.2012.2 and grant of Russian Ministry of
Education and Science, project TSPU-122. The work of K.V.S was
supported by RFBR grant, project No 14-01-00695.

\appendix

\bigskip
\bigskip

\noindent {\Large\bf Appendix}

\section{Higher derivative term invariant under ${\cal N}=2$ supersymmetry}
\hspace{\parindent}\label{Appendix_Noether}

In order to construct the action $S_\Lambda$, given by Eq.
(\ref{Higher_Derivative_Term}), it is convenient to use the
Noether method \cite{West:1990tg,Buchbinder:1998qv} writing the
supersymmetry transformations in terms of ${\cal N}=1$ superfields
\cite{Gates:1983nr}. As a starting point we consider the action

\begin{eqnarray}
&& S_0 = -\frac{1}{32 e_0^2\Lambda^2} \mbox{tr} \int d^4x\,\Bigg\{
\mbox{Re} \int d^2\theta\, (e^\Omega W^a e^{-\Omega}) \bar\nabla^2
\nabla^2 (e^\Omega W_a
e^{-\Omega})\nonumber\\
&& + \int d^4\theta\,(e^{-\Omega^+} \Phi^+ e^{\Omega^+})
\bar\nabla^2 \nabla^2 (e^\Omega \Phi e^{-\Omega})\Bigg\},
\end{eqnarray}

\noindent where $\Lambda$ is a regularization parameter. (Its
dimension is equal to the dimension of a mass.) In order to
construct an action invariant under the transformations
(\ref{Second_SUSY}) by the Noether method, at the first step we
calculate the variation of the action $S_0$. The result is given
by the following (non-vanishing) expression:

\begin{eqnarray}
&& \delta S_0 = -\frac{i}{32e_0^2\Lambda^2}\mbox{tr} \int
d^4x\,d^4\theta\,\Bigg\{-4\eta\, e^\Omega W^a
e^{-\Omega}\Big[e^{-\Omega^+}\Phi^+ e^{\Omega^+},
\nabla^2(e^\Omega W_a e^{-\Omega})\Big]\nonumber\\
&& - 4\eta^* \bar\nabla^2(e^{-\Omega^+} \bar W^{\dot a}
e^{\Omega^+}) \Big[ e^{-\Omega^+} \bar W_{\dot a} e^{\Omega^+},
e^{\Omega}\Phi e^{-\Omega}\Big] + \eta^* e^{-\Omega^+}\Phi^+
e^{\Omega^+} \Big[e^{\Omega}\Phi e^{-\Omega}, \bar\nabla^2
\nabla^2
(e^{\Omega}\Phi e^{-\Omega})\Big]\vphantom{\Bigg(}\nonumber\\
&&  - \eta^* e^{-\Omega^+}\Phi^+ e^{\Omega^+} \bar\nabla^2
\Big[e^{\Omega}\Phi e^{-\Omega}, \nabla^2 (e^{\Omega}\Phi
e^{-\Omega})\Big] +\eta e^{-\Omega^+}\Phi^+ e^{\Omega^+}
\bar\nabla^2 \Big[e^{-\Omega^+}\Phi^+ e^{\Omega^+},
\nabla^2(e^{\Omega}\phi
e^{-\Omega})\Big]\vphantom{\Bigg(}\nonumber\\
&&  - \eta e^{-\Omega^+}\Phi^+ e^{\Omega^+} \bar\nabla^2 \nabla^2
\Big[e^{-\Omega^+}\Phi^+ e^{\Omega^+}, e^{\Omega}\Phi
e^{-\Omega}\Big] \Bigg\}.
\end{eqnarray}

\noindent These terms can be canceled by adding

\begin{eqnarray}
&& S_1 = -\frac{1}{4 e_0^2\Lambda^2} \mbox{tr} \int d^4x\,
d^4\theta\,\Bigg( (e^\Omega W^a e^{-\Omega})
\Big[\nabla_a(e^\Omega \Phi e^{-\Omega}),(e^{-\Omega^+} \Phi^+
e^{\Omega^+})\Big]
\vphantom{\frac{1}{2}}\nonumber\\
&& + (e^{-\Omega^+} \bar W^{\dot a} e^{\Omega^+}) \Big[(e^\Omega
\Phi e^{-\Omega}),\bar\nabla_{\dot a}(e^{-\Omega^+} \Phi^+
e^{\Omega^+})\Big] \Bigg)
\end{eqnarray}

\noindent to the action $S_0$. The sum $S_0+ S_1$ is also not
invariant under the transformations (\ref{Second_SUSY}):

\begin{eqnarray}
&& \delta (S_0 + S_1) = -\frac{i}{2e_0^2\Lambda^2} \mbox{tr} \int
d^4x\,d^4\theta\, \Bigg(D_a\eta \Big[e^\Omega W^a e^{-\Omega},
e^{-\Omega^+}\Phi^+ e^{\Omega^+}\Big]\nonumber\\
&& + \bar D_{\dot a} \eta^+ \Big[e^{-\Omega^+}\bar W^{\dot a}
e^{\Omega^+}, e^\Omega\Phi e^{-\Omega}\Big]\Bigg) \Big[ (e^\Omega
\phi e^{-\Omega}), (e^{-\Omega^+} \Phi^+ e^{\Omega^+})\Big].
\end{eqnarray}

\noindent These terms can be canceled by adding the term

\begin{eqnarray}
&& S_2 = \frac{1}{2e_0^2\Lambda^2} \mbox{tr} \int d^4x\,
d^4\theta\,  \Big[ (e^\Omega \Phi e^{-\Omega}), (e^{-\Omega^+}
\Phi^+ e^{\Omega^+})\Big]^2
\end{eqnarray}

\noindent to the action. Then the sum

\begin{equation}
S_\Lambda = S_0 + S_1 + S_2
\end{equation}

\noindent is invariant under the transformations
(\ref{Second_SUSY}).


\end{document}